\documentclass[%
 reprint,nofootinbib
]{revtex4-2}

\usepackage{dcolumn}
\usepackage{physics,amsmath,amsthm,amsfonts,graphicx,times,xfrac,mathtools,amssymb,bbm,verbatim,appendix,mathrsfs,scalerel}
\usepackage[normalem]{ulem}
\usepackage{xcolor}
\theoremstyle{plain}
\newtheorem{definition}{Definition}
\newtheorem{proposition}{Proposition}
\newtheorem{theorem}{Theorem}
\newtheorem{remark}{Remark}
\newtheorem{example}{Example}

\newtheorem{manualpropositioninner}{Proposition}
\newenvironment{manualproposition}[1]{%
  \renewcommand\themanualpropositioninner{#1}%
  \manualpropositioninner
}{\endmanualpropositioninner}

\newtheorem{manualtheoreminner}{Theorem}
\newenvironment{manualtheorem}[1]{%
  \renewcommand\themanualtheoreminner{#1}%
  \manualtheoreminner
}{\endmanualtheoreminner}

\newcommand{\Def}{\coloneqq}
\newcommand{\Id}{\Tilde{\mathbb{I}}}
\newcommand{\rmin}{\mathrm{in}}
\newcommand{\rmout}{\mathrm{out}}
\newcommand{\rmi}{\mathrm{i}}
\newcommand{\rmo}{\mathrm{o}}
\newcommand{\rmA}{\mathrm{A}}
\newcommand{\rmB}{\mathrm{B}}
\newcommand{\rmAB}{\mathrm{AB}}
\newcommand{\rmC}{\mathrm{C}}
\newcommand{\rmD}{\mathrm{D}}
\newcommand{\rmE}{\mathrm{E}}
\newcommand{\rmI}{\mathrm{I}}
\newcommand{\rmM}{\mathrm{M}}
\newcommand{\rmN}{\mathrm{N}}
\newcommand{\rmR}{\mathrm{R}}
\newcommand{\Bcal}{\mathcal{B}}
\newcommand{\Ecal}{\mathcal{E}}
\newcommand{\Hcal}{\mathcal{H}}
\newcommand{\Ical}{\mathcal{I}}

\newcommand{\Mcal}{\mathcal{M}}

\newcommand{\Tt}{\mathcal{T}_2}
\newcommand{\Tn}{\mathcal{T}_n}

\begin{document}

\title{Relations between Markovian and non-Markovian correlations in multitime quantum processes}

\author{Guilherme Zambon}
\email{guilhermezambon@usp.br}
\author{Diogo O. Soares-Pinto}
\email{dosp@ifsc.usp.br}
\affiliation{
 Instituto de F{\'i}sica de S{\~a}o Carlos, Universidade de S{\~a}o Paulo, CP 369, 13560-970 S{\~a}o Carlos, SP, Brasil.
 }

\begin{abstract}
  In the dynamics of open quantum systems, information may propagate in time through either the system or the environment, giving rise to Markovian and non-Markovian temporal correlations, respectively. However, despite their notable coexistence in most physical situations, it is not yet clear how these two quantities may limit the existence of one another. Here, we address this issue by deriving several inequalities relating the temporal correlations of general multitime quantum processes. The dynamics are described by process tensors, and the correlations are quantified by the mutual information between subsystems of their Choi states. First, we prove a set of upper bounds to the non-Markovianity of a process given the degree of Markovianity in each of its steps. This immediately implies a nontrivial maximum value for the non-Markovianity of any process, independent of its Markovianity. Finally, we determine how the non-Markovianity limits the amount of total temporal correlations that could be present in a given process. These results show that, although any multitime process must pay a price in total correlations to have a given amount of non-Markovianity, this price vanishes exponentially with the number of steps of the process, while the maximum non-Markovianity grows only linearly. This implies that even a highly non-Markovian process might be arbitrarily close to having the maximum amount of total correlations if it has a sufficiently large number of steps.
\end{abstract}

\maketitle

\section{Introduction}

Understanding how information flows in quantum processes is a central task for developing quantum technologies and for better comprehending fundamental aspects of quantum theory. As any quantum system of interest inevitably interacts with an uncontrolled environment \cite{breuer2002theory,rivas2012open}, crucial information stored on the system may get lost during the dynamics \cite{nielsen2010quantum,wilde2013quantum}, which constitutes the main challenge for experimental implementation of quantum information processing protocols \cite{preskill2018quantum}.

Sometimes, however, this information returns. Such information backflow is what characterizes non-Markovian quantum processes \cite{rivas2014quantum,breuer2016colloquium,vega2017dynamics,li2018concepts}, whose dynamics are best described within the process tensor framework \cite{pollock2018non}. Unlike traditional approaches that typically employ quantum channels mapping an initial state to an evolved one \cite{breuer2002theory,rivas2012open,nielsen2010quantum,wilde2013quantum}, process tensors are generalizations of joint probabilities to the quantum realm \cite{milz2020kolmogorov}, which allows for a consistent definition of quantum Markovianity \cite{pollock2018operational} and proper treatment of memory effects \cite{milz2019completely,taranto2019quantum,taranto2019structure,figueroa2019almost,milz2020when,milz2021genuine,taranto2021non,figueroa2021markovianization,sakuldee2022connecting,capela2022quantum,taranto2023hidden,taranto2023characterising}.

In this way, several fields of quantum theory that had so far been explored only for Markovian dynamics are now being expanded to non-Markovian settings, which brings them closer to practical applications of the theory since, like in classical stochastic processes \cite{van1992stochastic,van1998remarks}, non-Markovianity is the rule in nature, not the exception. Examples of this may be found in the fields of quantum simulation \cite{jorgensen2019exploiting,jorgensen2020discrete,xiang2021quantify,cygorek2022simulation,fowler2022efficient,gribben2022using,cygorek2023sublinear,fux2023tensor}, randomized benchmarking \cite{figueroa2021randomized,figueroa2022towards,figueroa2023operational}, quantum process tomography \cite{milz2018reconstructing,white2020demonstration,white2021many,white2022non,white2022characterization,white2023filtering,aloisio2023sampling}, and quantum thermodynamics \cite{strasberg2019repeated,figueroa2020equilibration,huang2022fluctuation,huang2023multiple,dowling2023relaxation,dowling2023equilibration}, among others \cite{guo2020tensor,huang2023leggettgarg,butler2023optimizing}.

The main feature that allows for process tensors to best describe non-Markovian dynamics is their genuine multitime structure, which may be equivalently characterized by means of quantum combs \cite{chribella2008quantum} and quantum networks \cite{chiribella2009theoretical}. Similarly, one could also consider treating such correlations with slightly more general objects called process matrices, which are used in the context of quantum causal modeling because they can also describe processes with indefinite causal order \cite{costa2016quantum,costa2018unifying,milz2018entanglement,nery2021simple,giarmatzi2021witnessing,milz2022resource,milz2023transformations}. In this way, we could say that a process tensor is a time-ordered process matrix.\footnote{Although our approach is focused on process tensors, we indicate which of our results also hold for process matrices without a definite causal order.}

Importantly, it has been shown that one can construct resource theories of quantum processes which take process tensors as objects and superprocesses as transformations \cite{berk2021resource}. This is extremely useful, as it allows one to take relevant concepts, techniques, and even results from other resource theories and adapt them to the context of quantum processes. For example, Ref. \cite{berk2023extracting} proved that the non-Markovianity and time resolution of a process could be consumed as resources to reduce noise in a given process. This is done by means of an optimized dynamical decoupling protocol that takes the memory effects of the process into account.

\begin{figure}[t]
    \centering
    \includegraphics[width=\columnwidth]{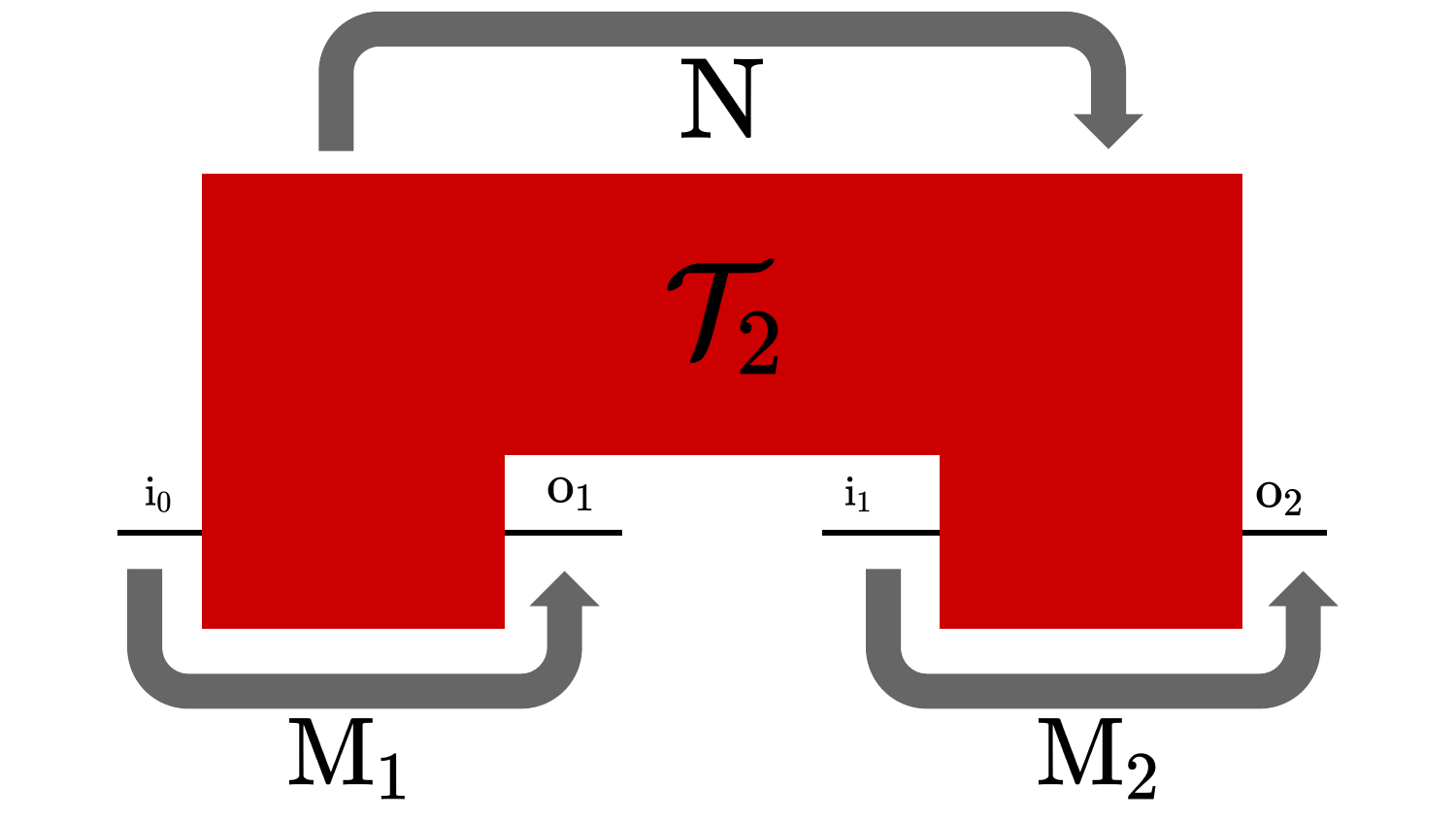}
    \caption{Markovian and non-Markovian correlations in a two-step quantum process. While $\rmM_j$ quantifies how much information is transmitted through the system in the $j$th step, $\rmN$ quantifies how much information is transmitted through the environment from the first step to the second one. In this two-step setting, we show that $\rmN\le 2\qty(2\ln d -\rmM_1)$ and $\rmN\le 2\ln d -\rmM_2$. The generalization of this result to the $n$-step scenario implies that the maximum non-Markovianity of an $n$-step process tensor is $2(n-1)\ln d$ and not $2n\ln d$, as it could seem at first, and also that the total correlations $\rmI$ of the process are upper bounded by $2n\ln d-\rmN/(2^n-2)$, meaning that a process can approach the maximum amount $2n\ln d$ of total correlations only if it has low non-Markovianity or a large number of steps.}
    \label{fig:correlations}
\end{figure}

As is usually the case for resource theories \cite{chitambar2019quantum}, monotones play a central role in the resource theory of quantum processes. Reference \cite{berk2023extracting} defined three important information quantifiers for general multitime processes, Markovian correlations, non-Markovian correlations and total correlations, and showed them to be monotonic under the free operations of the theory. Here, we take one step further in this direction and seek the relevant properties of these quantifiers. We use the notion of information exchange between the system and the environment and the time ordering of the process to prove several inequalities relating the multitime correlations of any given quantum process. A summary of our results is shown in Fig. \ref{fig:correlations}.

This paper is structured as follows. In Sec. \ref{sec:single-step} we show how the approach we use to quantify temporal correlations applies for single-step processes, as described by quantum channels. In Sec. \ref{sec:two-step} we present the two-step scenario, which is the simplest one in which non-Markovianity may take place, and discuss from an informational standpoint why Markovian and non-Markovian correlations should limit one another, proving some insightful bounds for this simple case. In Sec. \ref{sec:multitime} the previous analyses are put on firm mathematical grounds, and the results are generalized to the $n$-step scenario. The discussions are then concluded in Sec. \ref{sec:conclusion}.

\section{Temporal correlations in single-step quantum processes}\label{sec:single-step}

Quantum channels are the simplest descriptors of open quantum systems dynamics. They are linear, completely positive, and trace-preserving maps, which take an initial state of the system as input and provide as output the corresponding final state after a single interaction with the environment \cite{nielsen2010quantum,wilde2013quantum}. An example of the quantum channel which will be presented in several parts of this text is the depolarizing channel $\Ecal_{p}:\Bcal(\Hcal_{\rmin})\to\Bcal(\Hcal_{\rmout})$, whose action is defined as
\begin{equation}
\label{eq:dep}
    \Ecal_p(\rho) = p\Id+(1-p)\rho,
\end{equation}
where $\Id=\mathbb{I}/d$ is the maximally mixed state, $d$ is the dimension of the system and $p\in[0,1]$ is a parameter of the channel. For $p=0$ we have an identity channel $\Ecal_0(\rho) = \rho$, implying that in this case all the information about the initial state is preserved in the final state, such that the input and output are maximally correlated. On the other hand, for $p=1$ we have the completely depolarizing channel $\Ecal_1(\rho) = \Id$, which means that all the information about the initial state is lost during the process; i.e., the input and output are totally uncorrelated.

The amount of information that is preserved or lost in a quantum channel may be quantified in several ways. Here, we adopt an approach that has been shown to be well suited for multitime processes which consists of first mapping the temporal correlations of the process to spatial correlations of a corresponding state and then applying distance-based measures to quantify them \cite{luo2012quantifying,bylicka2014non,bylicka2016thermodynamic,pollock2018operational,berk2023extracting}.

The first step is achieved using the Choi-Jamio\l{}kowski isomorphism, in which a channel $\Ecal:\Bcal(\Hcal_{\rmin})\to\Bcal(\Hcal_{\rmout})$ is mapped to a state $\Upsilon^{\Ecal}\in\Bcal(\Hcal_{\rmin}\otimes\Hcal_{\rmout})$ by means of
\begin{equation}
    \Upsilon^{\Ecal} \Def \qty(\Ical_{\rmin}\otimes \Ecal)\Phi,
\end{equation}
where
\begin{equation}
    \Phi = \frac{1}{d}\sum_{i,j=1}^d\ket{i}\bra{j}_{\rmin}\otimes \ket{i}\bra{j}_\rmin
\end{equation}
is a normalized maximally entangled state in $\Bcal(\Hcal_{\rmin}\otimes\Hcal_{\rmin})$ and $\Ical_{\rmin}$ is the identity channel in $\Bcal(\Hcal_{\rmin})$.\footnote{Since we use the entropy of Choi states to quantify correlations, we opt for this less common definition in which they are normalized.} The trace preservation of $\Ecal$, which corresponds to its deterministic implementation, implies the trace condition
\begin{equation}
\label{eq:trace_cond}
    \tr_\rmout\qty[\Upsilon^{\Ecal}] = \Id_{\rmin}.
\end{equation}
In turn, any quantum state $\Upsilon\in\Bcal(\Hcal_{\rmin}\otimes\Hcal_{\rmout})$ satisfying the above condition may be associated with a quantum channel $\Ecal^{\Upsilon}:\Bcal(\Hcal_{\rmin})\to\Bcal(\Hcal_{\rmout})$ by
\begin{equation}
    \Ecal^{\Upsilon}(\rho_{\rmin}) = d\tr_\rmin\qty[\qty(\rho_{\rmin}\otimes \Id_{\rmout})\Upsilon^{T_\rmin}],
\end{equation}
where ${T_\rmin}$ represents the partial transpose in the $\ket{i}$ basis of $\Hcal_{\rmin}$.

Since the Choi state $\Upsilon^{\Ecal}$ contains all the information about the quantum channel $\Ecal$, the temporal correlations of the channel are mapped into spatial correlations between the subspaces $\Hcal_{\rmin}$ and $\Hcal_{\rmout}$ of $\Upsilon^{\Ecal}$. For example, 
\begin{equation}
    \Upsilon^{\Ecal}=\Tilde{\mathbb{I}}_{\rmin}\otimes \sigma_\rmout\iff\Ecal(\rho_{\rmin}) = \sigma_\rmout ~~~\forall \rho_{\rmin};
\end{equation}
that is, the Choi state is a product state iff the channel has a fixed output (i.e., having no spatial correlations is equivalent to having no temporal correlations). On the other hand, the Choi state is maximally entangled iff the channel is unitary, and it is separable iff the channel is entanglement-breaking \cite{verstraete2002quantum}.

\begin{figure}[t]
    \centering
    \includegraphics[width=\columnwidth]{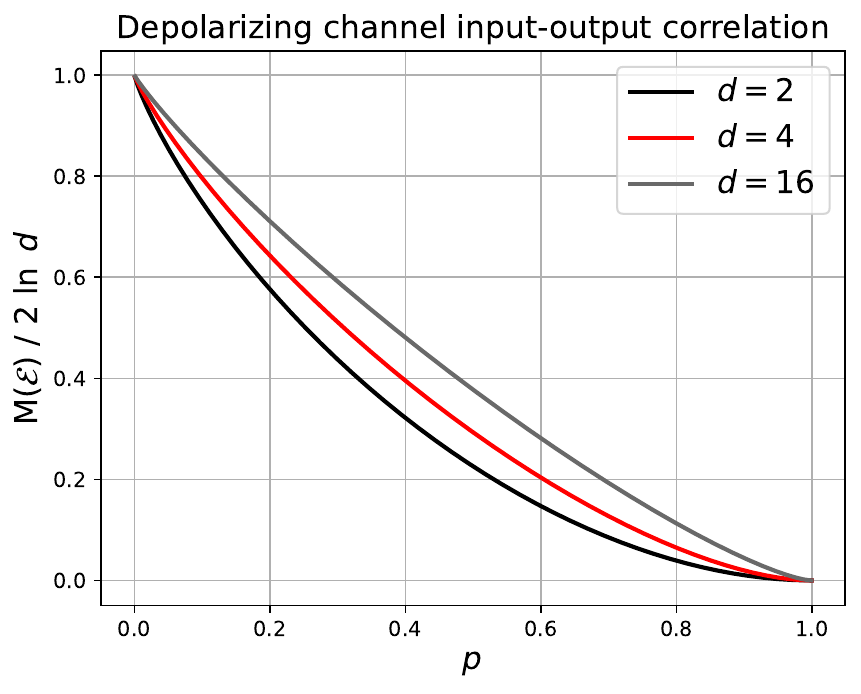}
    \caption{Input-output correlation of the depolarizing channel for different dimensions of the system. $\rmM(\Ecal_p)$ is a monotonic function of $p$ for any $d$. In general, the quantity $\rmM(\Ecal)$ ranges from $0$ for fixed output channels to $2\ln d$ for unitary channels.}
    \label{fig:qubit_dep}
\end{figure}

Now, to quantify the spatial correlations of Choi states we apply the mutual information between subspaces, which is equivalent to the relative entropy between the state and the closest uncorrelated state \cite{wilde2013quantum,berk2023extracting},
\begin{align}
\label{eq:mutual_relative}
    \rmI(\rmA:\rmB)_{\rho_\rmAB} &\Def S_\rmA+S_\rmB-S_\rmAB\\
    &= S(\rho_\rmAB||\rho_\rmA\otimes\rho_\rmB),
\end{align}
where $S_\rmA=-\tr[\rho_\rmA\ln \rho_\rmA]$ is the von Neumann entropy of $\rho_\rmA = \tr_\rmB\qty[\rho_\rmAB]$ and $S(\rho||\sigma)=\tr[\rho\qty(\ln \rho-\ln\sigma)]$ is the relative entropy quasi-distance between the states $\rho$ and $\sigma$. In this way, we can define the input-output correlation of the channel $\Ecal$ by
\begin{align}
    \rmM(\Ecal) &\Def \rmI(\rmin:\rmout)_{\Upsilon^\Ecal}\label{eq:def_M}\\
    &= S_{\rmin}+S_{\rmout}-S_{\rmin~\rmout}\label{eq:M_S}.
\end{align}
To illustrate how $\rmM$ quantifies the amount of information preserved by a given quantum channel, in Fig. \ref{fig:qubit_dep} we plot $\rmM(\Ecal_p)$ as a function of $p$ for the depolarizing channel.

\begin{figure}[t]
    \centering
    \includegraphics[width=\columnwidth]{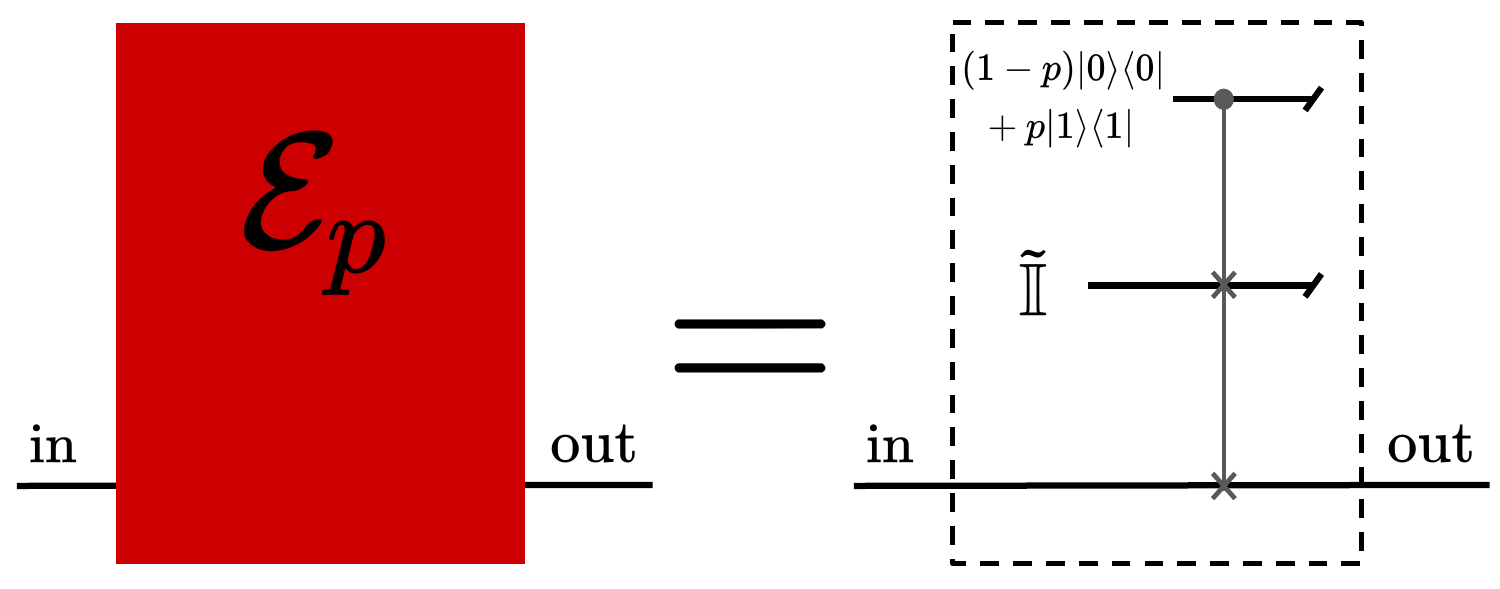}
    \caption{Example of a Stinespring dilation for the depolarizing channel in Eq. \eqref{eq:dep}. The global unitary is a controlled-SWAP (Fredkin) gate; the initial state of the control space of the environment is $(1-p)\ket{0}\bra{0}+p\ket{1}\bra{1}$, and the initial state of the target space of the environment is $\Id$. In this way, the input state $\rho$ of the system is swapped with $\Id$ conditioned on the state of the control space. Tracing out the environment yields a depolarizing reduced dynamics for the system.}
    \label{fig:stines_dep}
\end{figure}

To better understand the informational interplay between the system and the environment, consider the Stinespring dilation theorem \cite{stinespring1955positive,nielsen2010quantum,wilde2013quantum}, according to which the action of any given channel $\Ecal$ may be represented as 
\begin{equation}
\label{eq:stinespring}
    \Ecal(\rho) = \tr_\rmE\qty[U(\rho\otimes\sigma_\rmE)U^\dagger] 
\end{equation}
for some initial state $\sigma_\rmE\in\Bcal(\Hcal_\rmE)$ of the environment and some global unitary $U:\Bcal(\Hcal_\rmin\otimes\Hcal_\rmE)\to\Bcal(\Hcal_\rmout\otimes\Hcal_\rmE)$. Figure \ref{fig:stines_dep} shows a possible dilation for the depolarizing channel.

Now, consider a purification $\Psi_{\mathrm{ER}}=\ket{\psi}\bra{\psi}\in\Bcal(\Hcal_\rmE\otimes\Hcal_\rmR)$ of $\sigma_\rmE$ to an ancilla $\rmR$, that is,
\begin{equation}
    \tr_{\scaleto{\rmR}{5pt}}\qty[\Psi_{\mathrm{ER}}]=\sigma_\mathrm{E},
\end{equation}
and define a global pure state $\eta\in\Bcal(\Hcal_\rmin\otimes\Hcal_\rmout\otimes\Hcal_\rmE\otimes\Hcal_\rmR)$ after the unitary interaction
\begin{equation}
    \eta \Def \qty(\mathcal{I}_{\mathrm{in}}\otimes \mathcal{U}^{\rmin,\rmE}_{\rmout,\rmE}\otimes \mathcal{I}_\mathrm{R})\qty(\Phi\otimes \Psi_{\mathrm{ER}}).
\end{equation}
Comparing this definition to Eqs. \eqref{eq:def_M} and \eqref{eq:stinespring}, it is immediately clear that
\begin{equation}
    \rmI(\rmin:\rmout)_{\eta} = \rmM.
\end{equation}
By defining the complement of $\rmM$ as
\begin{equation}
    \overline{\rmM}\Def 2\ln d-\rmM,
\end{equation}
it follows that
\begin{align}
    \rmI(\rmin:\rmE\rmR)_{\eta} &= S_\rmin+S_{\rmE\rmR}-S_{\rmin~\rmE\rmR}\\
    &= S_\rmin+S_{\rmin~\rmout}-S_{\rmout}\\
    &= 2S_\rmin-(S_\rmin+S_{\rmout}-S_{\rmin~\rmout})\\
    &= 2\ln d - \rmM\\
    &=\overline{\rmM} \label{eq:in_E_R},
\end{align}
where we used $S_{\rmE\rmR}=S_{\rmin~\rmout}$ and $S_{\rmin~\rmE\rmR}=S_{\rmout}$ since bipartitions of pure states share the same entropy \cite{nielsen2010quantum,wilde2013quantum} and also $S_\rmin=\ln d$ because, from Eq. \eqref{eq:trace_cond}, we know $\tr_{\scaleto{\rmout\rmE\rmR}{5pt}}\qty[\eta]$ is maximally mixed. Equation \eqref{eq:in_E_R} shows that, while $\rmM$ is the information about the initial state of the system that is kept in its the final state, $\overline{\rmM}$ is exactly the part of this information that is lost to the environment. In turn, the system might also get some information about the initial state of the environment. Since this information is useless to us, it is simply treated as noise. However, that will not always be the case in multitime processes, as at any time step that is not the first one the environment may be carrying information about past states of the system.

Importantly, we can show that for the system to get this information from the environment, it must also give some of its own. To see this, we calculate the mutual information between the system and the ancilla $\mathrm{R}$. Since $\mathrm{R}$ is initially entangled with the environment $\mathrm{E}$ and is not affected by the interaction, it contains information about the initial state of $\mathrm{E}$. Notice that
\begin{align}
    \rmI(\rmin~\rmout:\rmR)_{\eta} &= S_{\rmin~\rmout}+S_{\rmR}-S_{\rmin~\rmout~\rmR}\label{eq:exchange_i}\\
    &\le 2S_{\rmin~\rmout}\\
    &= 2(S_\rmin+S_{\rmout}-\rmM)\\
    &\le 2(2\ln d - \rmM)\\
    &=2\overline{\rmM},\label{eq:exchange_f}
\end{align}
where the relation $S_{\rmR}-S_{\rmin~\rmout~\rmR}\le S_{\rmin~\rmout}$ is a consequence of Araki-Lieb triangle inequality \cite{araki1970entropy} 
\begin{equation}
    \label{eq:AL}
    |S_\rmA-S_\rmB|\le S_\rmAB
\end{equation}
and we also used $S_\rmin+S_{\rmout}\le2\ln d$. Equation \eqref{eq:exchange_f} means the amount of information the system might receive about the initial state of the environment is upper bounded by twice the amount of information the environment receives about the initial state of the system. This intuition on information exchange between the system and the environment lies at the heart of our relations between temporal correlations in multitime processes.

\section{Temporal correlations in two-step quantum processes}\label{sec:two-step}

\begin{figure}[t]
    \centering
    \includegraphics[width=\columnwidth]{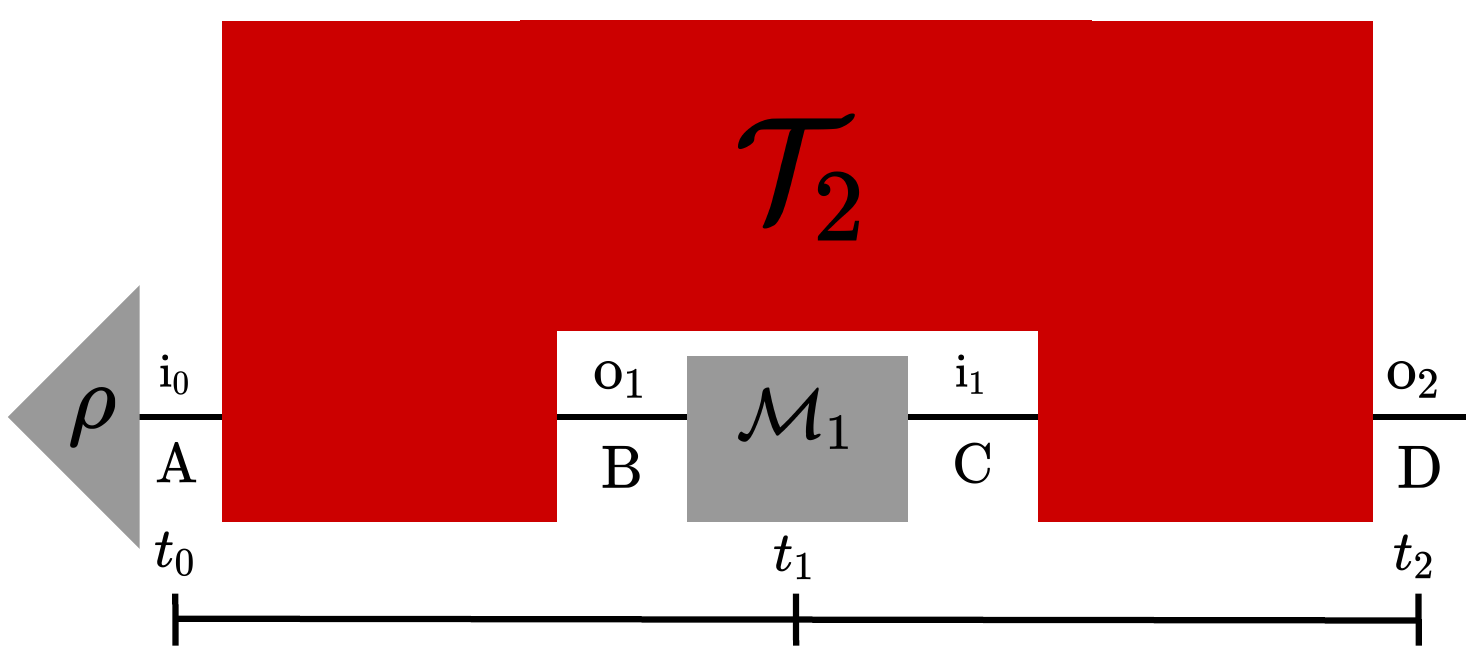}
    \caption{The two-step scenario. Initially, an experimenter prepares the system in a state $\rho$ as the input at time $t_0$. Then, the system interacts with the environment in a first step, resulting in a new output state at time $t_1$. Next, the experimenter performs a control operation on the system alone, as described by a quantum channel $\Mcal_1:\Bcal(\Hcal_{\rmo_1})\to\Bcal(\Hcal_{\rmi_1})$. Finally, still at time $t_1$, the system interacts in a second step with the same environment as before, yielding a final state at time $t_2$. The Hilbert space labels above the lines describe whether the corresponding state at a given time is the input or the output of an interaction with the environment, such that $\rmo_1$ is the output space at time $t_1$, for example. As a consequence of this convention, the experimenter operations will be mappings from output spaces to input spaces at fixed times. To simplify the notation, for the remainder of this section we adopt the Hilbert space labeling shown below the lines; that is, we use $\{\rmA,\rmB,\rmC,\rmD\}$ instead of $\{\rmi_0,\rmo_1,\rmi_1,\rmo_2\}$. This more descriptive labeling will be recovered in the discussion of the multitime scenario in Sec. \ref{sec:multitime}.}
    \label{fig:two_step}
\end{figure}

Besides the single-step processes described by quantum channels, one could also consider a two-step scenario, as depicted in Fig. \ref{fig:two_step}. Notice that in this case the action of the environment on the system will not, in general, be described by a quantum channel in each step, as the global state before the second step might not be a product one and, even if it is, the state of the environment might be conditioned on the input state at time $t_0$. Thus, the most general descriptor of the action of the environment in the two-step scenario will be a mapping $\Tt$ taking both the initial state and the control operation to the final state,
\begin{equation}
    \Tt\qty[\rho\otimes\Mcal_1] = \rho^\prime.
\end{equation}
To preserve the physical properties of the process, this mapping must be multi-linear, completely positive, trace preserving, and time ordered. A mapping $\Tt:\Bcal(\Hcal_{\rmA})\otimes \Bcal^2(\Hcal_{\rmB},\Hcal_{\rmC})\to\Bcal(\Hcal_{\rmD})$ satisfying these conditions is what we call a two-step process tensor \cite{pollock2018non}.

\begin{figure}[t]
    \centering
    \includegraphics[width=\columnwidth]{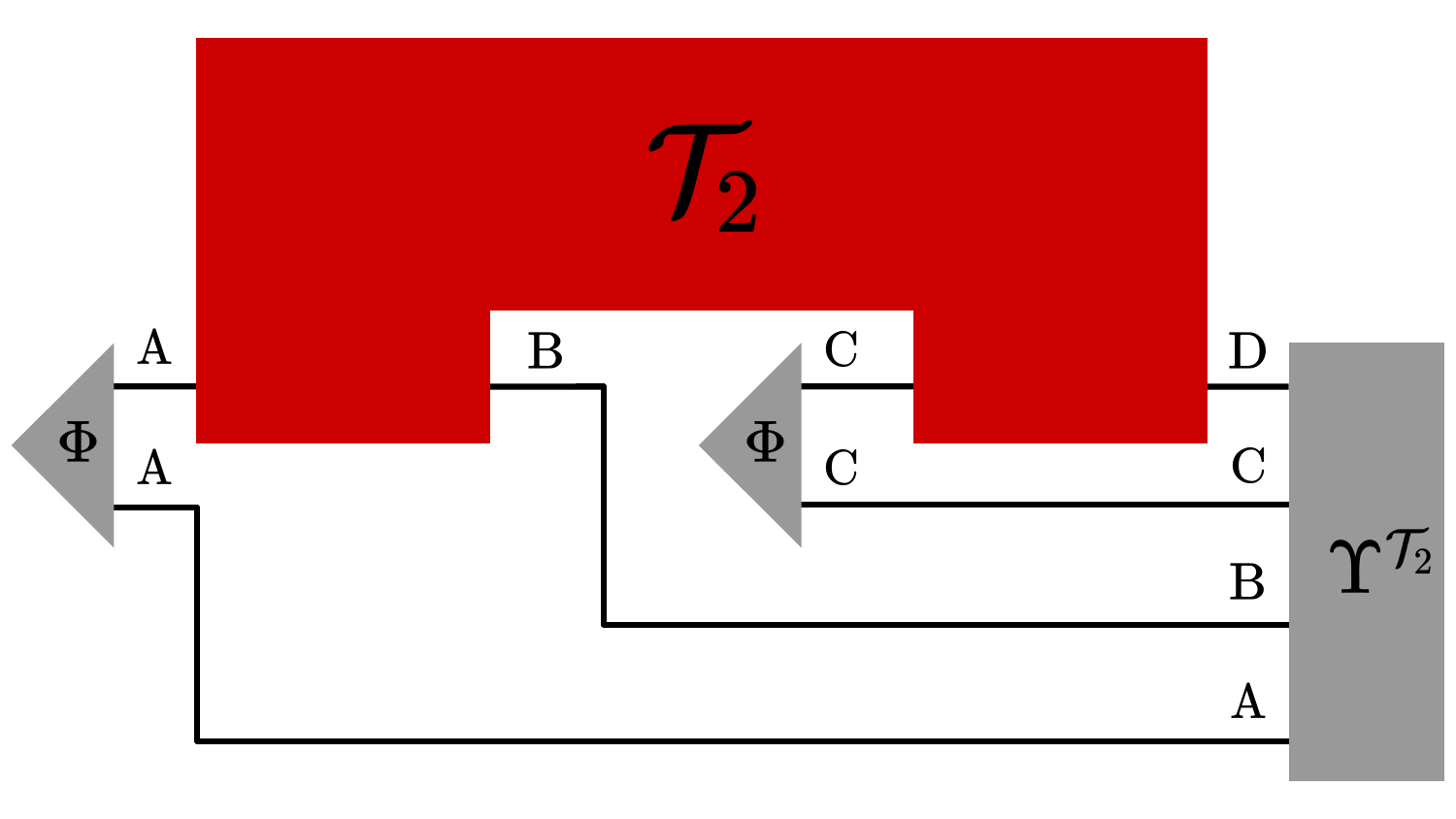}
    \caption{Circuit generating the Choi state of a given process tensor $\Tt$. At each time step we prepare a maximally entangled state $\Phi$ and store half of the state while letting the other half interact with the environment. The state of the system after each interaction is then also stored. The global quadripartite state stored in the end is the Choi state of the process tensor describing this two-step process.}
    \label{fig:choi_two}
\end{figure}

Alternatively, $\Tt$ might be seen as a mapping from two input spaces to two output spaces, $\Tt:\Bcal(\Hcal_{\rmA}\otimes\Hcal_{\rmB})\to\Bcal(\Hcal_{\rmC}\otimes\Hcal_{\rmD})$. In this way, we define the Choi state $\Upsilon^{\Tt}\in\Bcal(\Hcal_{\rmA}\otimes\Hcal_{\rmB}\otimes\Hcal_{\rmC}\otimes\Hcal_{\rmD})$ of the process tensor $\Tt$ as
\begin{equation}
\label{eq:choi_tt}
    \Upsilon^{\Tt}\Def \qty(\Ical_{\rmA}\otimes\Ical_{\rmC}\otimes\Tt)\qty(\Phi_{\rmA\rmA}\otimes\Phi_{\rmC\rmC}),
\end{equation}
as shown in Fig. \ref{fig:choi_two}. In terms of its Choi state, the trace-preservation and time ordering properties of the two-step process tensor $\Tt$ are given by
\begin{align}
    &\tr_{\scaleto{\rmD}{5pt}}\qty[\Upsilon^{\Tt}] = \tr_{\scaleto{\rmC\rmD}{5pt}}\qty[\Upsilon^{\Tt}]\otimes \Id_{\scaleto{\rmC}{5pt}}, \label{eq:order1} \\
    &\tr_{\scaleto{\rmB\rmC\rmD}{5pt}}\qty[\Upsilon^{\Tt}] = \Id_{\scaleto{\rmA}{5pt}}. \label{eq:order2}
\end{align}

Like in the single-step scenario, the temporal correlations of the two-step process are mapped into spatial correlations of its Choi state, with the additional feature that in this case there is more than just an input-output correlation. We follow Ref. \cite{berk2023extracting} and define the total correlations of the process $\Tt$
\begin{align}
\label{eq:def_I}
    \rmI(\Tt)\Def&~ \rmI(\rmA:\rmB:\rmC:\rmD)_{\Upsilon^{\Tt}}\\
    =&~ S_\rmA+ S_\rmB+ S_\rmC+ S_\rmD - S_{\rmA\rmB\rmC\rmD},
\end{align}
the Markovian correlations in the first step
\begin{align}
\label{eq:def_M1}
    \rmM_1(\Tt)\Def&~ \rmI(\rmA:\rmB)_{\Upsilon^{\Tt}}\\
    =&~ S_\rmA+ S_\rmB-S_{\rmA\rmB},
\end{align}
the Markovian correlations in the second step
\begin{align}
\label{eq:def_M2}
    \rmM_2(\Tt)\Def&~ \rmI(\rmC:\rmD)_{\Upsilon^{\Tt}}\\
    =&~ S_\rmC+ S_\rmD - S_{\rmC\rmD},
\end{align}
the total Markovian correlations
\begin{align}
\label{eq:def_Mkv}
    \rmM\Def&~ \rmM_1+\rmM_2,
\end{align}
and the non-Markovian correlations
\begin{align}
\label{eq:def_N}
    \rmN(\Tt)\Def&~ \rmI(\rmA\rmB:\rmC\rmD)_{\Upsilon^{\Tt}}\\
    =&~ S_{\rmA\rmB} + S_{\rmC\rmD}-S_{\rmA\rmB\rmC\rmD},
\end{align}
immediately implying the additive property
\begin{align}
\label{eq:add}
    \rmI &= \rmM+\rmN\\
    &= \rmM_1+\rmM_2+\rmN
\end{align}
for any two-step process $\Tt$.

\begin{figure}[t]
    \centering
    \includegraphics[width=\columnwidth]{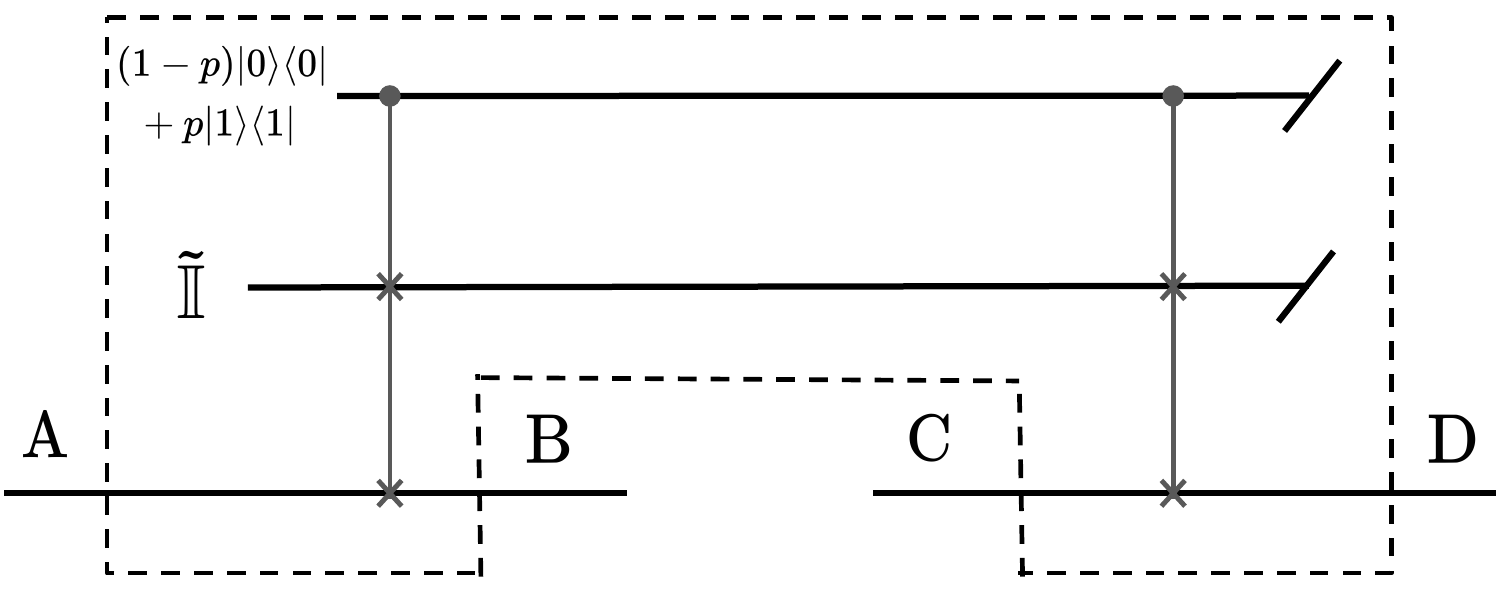}
    \caption{Two-step non-Markovian depolarizing dynamics. The first step of this dynamics is equal to the single-step process described in Fig. \ref{fig:stines_dep}, where depending on the state of the control system the input state is exchanged with a maximally mixed state. Then, in the second step we have the same unitary acting on the same systems, implying the second input is also exchanged with the state of the environment conditioned on the control system. Importantly, in this second step the environment may be carrying information about the first input, which could give rise to non-Markovian correlations in the dynamics.}
    \label{fig:nm_dep}
\end{figure}

\begin{figure}[t]
    \centering
    \includegraphics[width=\columnwidth]{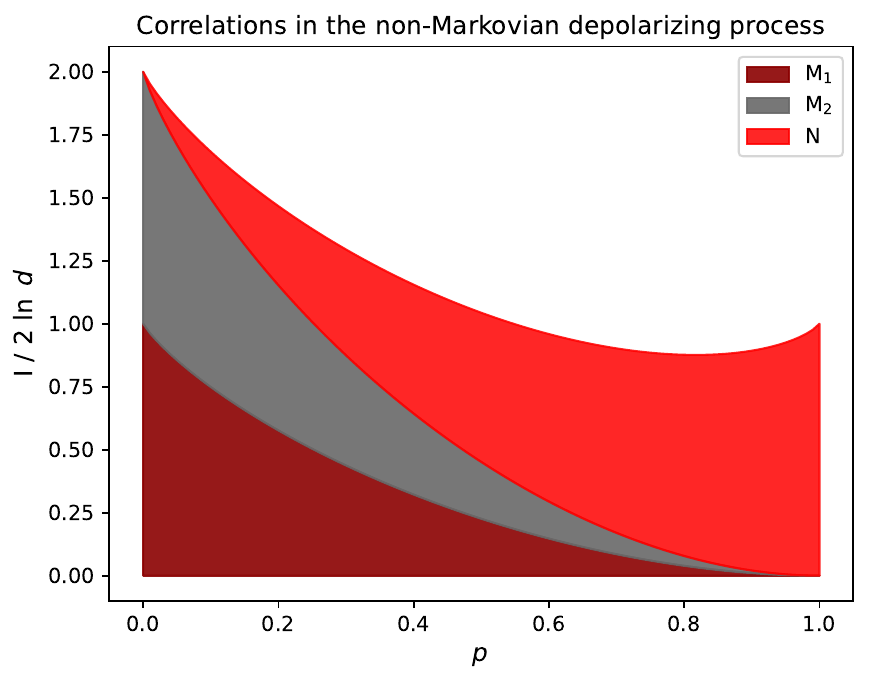}
    \caption{Correlations in the process shown in Fig. \ref{fig:nm_dep}. Specifically, in this process we always have $\rmM_1=\rmM_2$. For $p=0$ we have identity channels in both steps, implying $\rmM_1=\rmM_2=2\ln d$ and $\rmN=0$, with $\rmI=\rmM=4\ln d$. For $p>0$ there is information exchanged with the environment in both steps, such that $\rmM<4\ln d$, but part of the information lost in the first step is recovered in the second one, meaning $\rmN>0$. For $p=1$ if we input states $\rho_1$ and $\rho_2$ in the first and second steps, we obtain, respectively, $\Id$ and $\rho_1$ as outputs, such that $\rmM_1=\rmM_2=0$, as each output is uncorrelated with its input. However, since the second output is maximally correlated with the first input, we still have information propagating in time through the environment, implying this is a highly non-Markovian process, with $\rmI=\rmN=2\ln d$.}
    \label{fig:nm_dep_graph}
\end{figure}

Notice that $\rmM_1$ is the input-output correlation, as defined in Eq. \eqref{eq:def_M}, for the channel describing the first step of the process $\Tt$, which is associated with the Choi state $\tr_{\scaleto{\rmC\rmD}{5pt}}[\Upsilon^{\Tt}]$. Moreover, we can obtain a channel in the second step by averaging over all possible inputs in the first step of $\Tt$. Such a channel, whose Choi state is $\tr_{\scaleto{\rmA\rmB}{5pt}}[\Upsilon^{\Tt}]$, would have an input-output correlation equal to $\rmM_2$\footnote{The reason for calling such correlations Markovian is the operational character of the process tensor approach, in which we consider the experimenter to have access only to the system and only in the indicated discrete set of times \cite{pollock2018non}. Nothing is assumed about what happens in between these times or about the underlying global closed dynamics from which the observed system evolution could be arising. In this sense, even if the dynamics between two consecutive time steps was emerging from the discretization of a non-Markovian dynamics, such non-Markovianity could not be detected by our constrained experimenter. Thus, non-Markovianity here refers only to the memory effects that could be operationally perceived by the experimenter. Since all the other time correlations of the process could, in principle, be reproduced by Markovian models, they are regarded as Markovian \cite{pollock2018operational}.}. Finally, although we could see that $\rmN$ is a measure of correlations between the two steps, it might not be fully clear why it should be associated with the non-Markovianity of the process. To clarify this, we recall the operational Markov condition given in Ref. \cite{pollock2018operational}, according to which a process is Markovian if its Choi state is of the form
\begin{equation}
\label{eq:Markov-cond}
    \Upsilon_{\rmA\rmB\rmC\rmD} = \Upsilon_{\rmA\rmB}\otimes\Upsilon_{\rmC\rmD},
\end{equation}
implying the distance between any given Choi state and the closest Markov Choi state will measure the degree of non-Markovianity of the process. In this way, since the $\rmN$ we defined in Eq. \eqref{eq:def_N} is the relative entropy between $\Upsilon^{\Tt}$ and the closest Markov Choi state, it will be a non-Markovianity quantifier following the recipe of Ref. \cite{pollock2018operational}. A schematic representation of the Markovian and non-Markovian correlations in a two-step quantum process is shown in Fig. \ref{fig:correlations}. As an example of how these correlations may coexist in a physical situation, we consider the two-step non-Markovian depolarizing dynamics\footnote{We call this dynamics ``non-Markovian depolarizing'' because although each step is locally depolarizing, in the sense that $\Upsilon_{\rmA\rmB}$ and $\Upsilon_{\rmC\rmD}$ are Choi states of depolarizing channels, it is also non-Markovian, as its Choi state is not of the product form of Eq. \eqref{eq:Markov-cond}. Importantly, however, this process could never arise from a time-continuous depolarizing dynamics, as discretizations of CP-divisible dynamics can only yield Markovian processes, with Choi states of the form of Eq. \eqref{eq:Markov-cond}.} shown in Fig. \ref{fig:nm_dep}. In Fig. \ref{fig:nm_dep_graph} we see how they vary as a function of the parameter $p$.  

From the definitions of Eqs. \eqref{eq:def_I} to \eqref{eq:def_N} and the discussion in Sec. \ref{sec:single-step} we can see that
\begin{align}
    &0\le\rmI\le 4\ln d,\\
    &0\le\rmM_1\le 2\ln d,\\
    &0\le\rmM_2\le 2\ln d,\\
    &0\le\rmM\le 4\ln d,\\
    &0\le\rmN\le 4\ln d.
\end{align}
The saturation of the upper bound of the four first quantities above is achieved, for example, in the dynamics of Figs. \ref{fig:nm_dep} and \ref{fig:nm_dep_graph} for $p=0$. The maximum possible value for $\rmN$ will be later discussed in more detail. Notice that the additive property of Eq. \eqref{eq:add} implies
\begin{equation}
\label{eq:weak_bound}
    \rmM_1+\rmM_2+\rmN\le 4\ln d.
\end{equation}
This means that Markovian and non-Markovian correlations limit the existence of one another, so that they cannot vary independently within their ranges. In particular, if $\rmM_1=\rmM_2=2\ln d$ we must then necessarily have $\rmN=0$, an example of which we see in Fig. \ref{fig:nm_dep_graph}.

Nevertheless, Eq. \eqref{eq:weak_bound} does not carry all the physical properties of information exchange we expect the process to have. Consider, for example, a scenario in which $\rmM_1=2\ln d$ and $\rmM_2=0$. From Eq. \eqref{eq:weak_bound} we have $\rmN\le 2\ln d$, so we could say that it is still possible for the process to have some non-Markovianity. However, notice that $\rmM_1=2\ln d$ implies that the first step of the dynamics is a unitary on the system alone, so the system does not exchange with the environment any information that could be recovered in the second step, and we must necessarily have $\rmN=0$. Analogously, in an opposite scenario where $\rmM_1=0$ and $\rmM_2=2\ln d$, the system does not exchange information with the environment in the second step, so it cannot recover any information that was lost in the first one, also implying $\rmN=0$. This means that both $\rmM_1$ and $\rmM_2$ should somehow individually provide upper bounds to $\rmN$.

To obtain such bounds we proceed as in Eqs. \eqref{eq:exchange_i} to \eqref{eq:exchange_f}, that is,
\begin{align}
    \rmN &= S_{\rmA\rmB} + S_{\rmC\rmD}-S_{\rmA\rmB\rmC\rmD}\\
    &\le 2S_{\rmA\rmB}\\
    &= 2\qty(S_{\rmA}+S_{\rmB}-\rmM_1)\\
    &\le 2\qty(2\ln d-\rmM_1)
\end{align}
and also
\begin{align}
    \rmN &= S_{\rmA\rmB} + S_{\rmC\rmD}-S_{\rmA\rmB\rmC\rmD}\\
    &\le 2S_{\rmC\rmD}\\
    &\le 2\qty(2\ln d-\rmM_2).
\end{align}
In this way, after defining the complements of the Markovian correlations $\overline{\rmM}_j\Def 2\ln d - \rmM_j$, we obtain a set of conditions describing how the Markovianity on each step individually limits the non-Markovianity of the process,
\begin{equation}
\label{eq:cond1}
    \rmN\le
    \begin{cases}
        &2\overline{\rmM}_1,\\
        &2\overline{\rmM}_2.
    \end{cases}
\end{equation}
By adding these conditions and dividing by 2 we recover Eq. \eqref{eq:weak_bound}. However, these more detailed bounds go beyond Eq. \eqref{eq:weak_bound} in that they show how the presence of non-Markovian correlations require information exchange between the system and the environment in both steps.

Nevertheless, the conditions in Eq. \eqref{eq:cond1} are still not the end of the story. That is because the time ordering condition in Eq. \eqref{eq:order1} was not used in their derivation; therefore, Eq. \eqref{eq:cond1} is valid even for process matrices without a definite causal order. Notice that Eq. \eqref{eq:order1} implies $S_{\rmA\rmB\rmC} = S_{\rmA\rmB} + \ln d$, so we have
\begin{align}
    \rmN &= S_{\rmA\rmB} + S_{\rmC\rmD}-S_{\rmA\rmB\rmC\rmD}\\
    &= S_{\rmA\rmB\rmC} -\ln d + S_{\rmC\rmD}-S_{\rmA\rmB\rmC\rmD}\\
    &\le S_{\rmD}-\ln d + S_{\rmC\rmD}\\
    &\le S_{\rmC\rmD}\\
    &\le \overline{\rmM}_2,
\end{align}
where we used $S_{\rmA\rmB\rmC} -S_{\rmA\rmB\rmC\rmD}\le S_{\rmD}$, implied by Eq. \eqref{eq:AL}$, S_{\rmD}\le \ln d$ and $S_{\rmC\rmD}\le 2\ln d - \rmM_2$. The set of conditions is then updated to
\begin{equation}
\label{eq:cond2}
    \rmN\le
    \begin{cases}
        &2\overline{\rmM}_1,\\
        &~~\overline{\rmM}_2.
    \end{cases}
\end{equation}

\begin{figure}[t]
    \centering
    \includegraphics[width=\columnwidth]{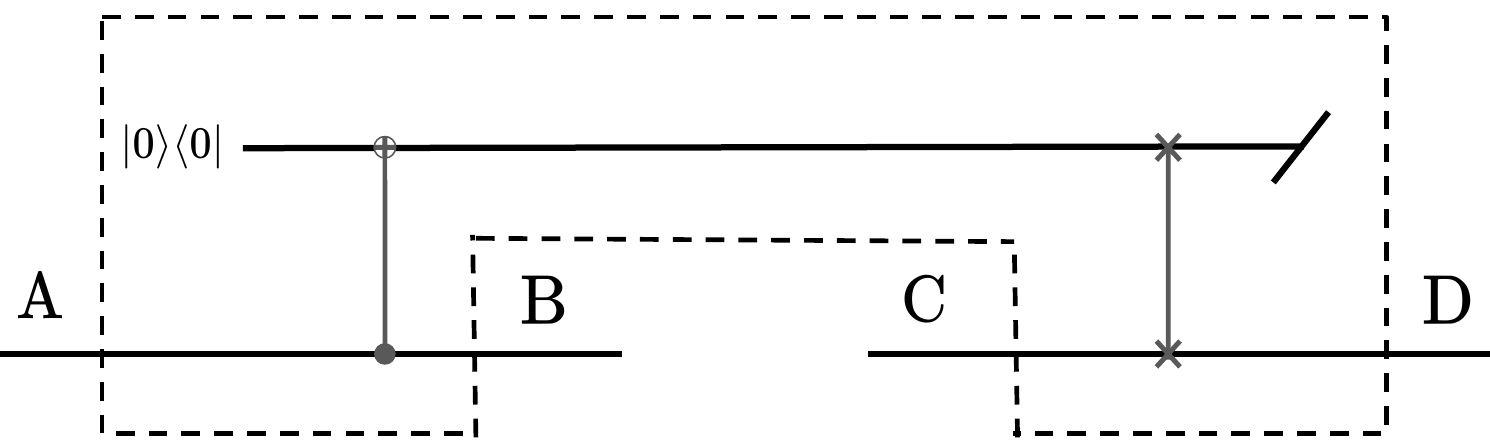}
    \caption{Maximally non-Markovian process with Markovian correlations in the first step. The first interaction is a CNOT with control on the system and target on the environment initially in the pure state $\ket{0}$. The second step is a SWAP between the system and the environment. The Choi state for this dynamics is $\Phi_{\rmA\rmB\rmD}\otimes\Id_\rmC$, where $\Phi_{\rmA\rmB\rmD}$ is a maximally entangled tripartite state, which means we have $\rmM_1=\ln d$, $\rmM_2=0$ and $\rmN=2\ln d$. Upon inputting states $\rho_1$ and $\rho_2$ into this process, we have state $\rho_1$ teleported to the Bell basis of the joint output space $\rmB\rmD$. Considering each output individually, we have two classical copies of $\rho_1$, that is, $\rho_1$ dephased in the computational basis.}
    \label{fig:nm_cnot}
\end{figure}

Such limitations to the non-Markovianity of any two-step process have several important implications. First, notice that $\rmN\le 2\ln d - \rmM_2\le 2\ln d$, so we can rewrite the range of $\rmN$ as
\begin{equation}
    0\le \rmN \le 2\ln d.
\end{equation}
This means that, despite there being states in $\Bcal(\Hcal_{\rmA}\otimes\Hcal_{\rmB}\otimes\Hcal_{\rmC}\otimes\Hcal_{\rmD})$ for which $2\ln d<\rmI(\rmA\rmB:\rmC\rmD)\le 4\ln d$, they are not Choi states of two-step process tensors, as they do not satisfy the time ordering conditions of Eqs. \eqref{eq:order1} and \eqref{eq:order2}. When restricted to the set of states that do satisfy such conditions, the maximum possible value for $\rmI(\rmA\rmB:\rmC\rmD)$ is actually $2\ln d$. An example of a process with this maximum non-Markovianity is that in Fig. \ref{fig:nm_dep} with $p=1$, where $\rmM_1=\rmM_2=0$. Importantly, while the condition $\rmM_2=0$ is necessary for $\rmN=2\ln d$, we could still have $0<\rmM_1\le \ln d$, an example of which is shown in Fig. \ref{fig:nm_cnot}.

More generally, we could say that a process that has high Markovianity in some step has necessarily low non-Markovianity, that is,
\begin{align}
    &\rmM_1\ge 2\ln d-\varepsilon \implies \rmN\le 2\varepsilon,\\
    &\rmM_2\ge 2\ln d-\varepsilon \implies \rmN\le \varepsilon.
\end{align}
On the other hand, a highly non-Markovian process is necessarily low in the Markovianity of its second step, but could still have medium Markovianity in the first one,
\begin{equation}
    \rmN\ge2\ln d-2\varepsilon \implies
    \begin{cases}
        &\rmM_1\le \ln d+\varepsilon,\\
        &\rmM_2\le 2\varepsilon.
    \end{cases}
\end{equation}

Another relevant feature brought by Eq. \eqref{eq:cond2} is that the interplay between Markovian and non-Markovian correlations is not symmetric. To see this, we rewrite the conditions in Eq. \eqref{eq:cond2} as $\rmM_1\le 2\ln d - \rmN/2$ and $\rmM_2\le 2\ln d - \rmN$, which implies
\begin{equation}
    \rmM \le 4\ln d -\frac{3}{2}\rmN;
\end{equation}
that is, for a process to have a given amount $\rmN$ of non-Markovianity it must ``pay a price'' in Markovianity of at least $3\rmN/2$. Alternatively, in terms of the total correlations of the process, we may write
\begin{equation}
    \rmI \le 4\ln d -\frac{1}{2}\rmN.
\end{equation}
This means that the only two-step processes with maximum total correlations are those with local unitaries on the system in both steps, so that they have maximum Markovian correlations and zero non-Markovian correlations, like in the example in Fig. \ref{fig:nm_dep} with $p=0$. On the other hand, a process with maximum non-Markovianity has at most $3\ln d$ total correlations, as in the example in Fig. \ref{fig:nm_cnot}. More generally, for a two-step process to approach the upper bound of $4\ln d$ on total correlations it must have as little non-Markovianity as possible,
\begin{equation}
    \rmI \ge 4\ln d-\varepsilon \implies \rmN\le 2\varepsilon.
\end{equation}
In contrast, any highly non-Markovian process has its total correlations limited to about $3\ln d$,
\begin{equation}
    \rmN \ge 2\ln d-2\varepsilon \implies \rmI\le 3\ln d +\varepsilon.
\end{equation}

In summary, we have discussed how the physical properties of information exchange and time ordering of the processes give rise to fundamental relations between the Markovian and non-Markovian correlations of any two-step quantum process. The mathematical constraints imposed by these properties allowed us to derive the conditions of Eq. \eqref{eq:cond2}. They, in turn, provided a general upper bound to non-Markovian correlations and revealed that non-Markovianity is undesirable if one seeks a two-step process with a large amount of total correlations. Now, we move to Sec. \ref{sec:multitime}, where all these relevant results are generalized to the $n$-step scenario.

\section{Temporal correlations in general multitime quantum processes}\label{sec:multitime}

We begin by defining the descriptor of the dynamics and its Choi state. We assume the most general multitime quantum process is described by an $n$-step process tensor, defined as follows.

\begin{definition}
    An $n$-step process tensor $\Tn:\Bcal(\Hcal_{\rmi_0})\otimes \Bcal^2(\Hcal_{\rmo_1},\Hcal_{\rmi_1})\otimes \cdots\otimes\Bcal^2(\Hcal_{\rmo_{n-1}},\Hcal_{\rmi_{n-1}})\to\Bcal(\Hcal_{\rmo_n})$ is a multi-linear, completely positive, trace-preserving and time-ordered mapping from an input state $\rho\in\Bcal(\Hcal_{\rmi_0})$ and a sequence $\{\Mcal_j\}_{j=1}^{n-1}$ of control operations $\Mcal_j:\Bcal(\Hcal_{\rmo_j})\to\Bcal(\Hcal_{\rmi_j})$ to a final state $\rho^\prime\in\Bcal(\Hcal_{\rmo_n})$.
\end{definition}

Notably, it is also possible to use a process tensor to describe a dynamics in which the system and the environment are initially correlated. In this case, $\Tn$ will be a mapping from the control operations to the final state of the system. Despite these slightly different conventions, the results we derive should also hold in those cases with small adaptations.

Importantly, the process tensor $\Tn$ may be equivalently seen as a time-ordered mapping from a sequence of inputs to a sequence of outputs, that is, $\Tn:\Bcal(\Hcal_{\rmi_0}\otimes\cdots\otimes\Hcal_{\rmi_{n-1}})\to\Bcal(\Hcal_{\rmo_1}\otimes\cdots\otimes\Hcal_{\rmo_{n}})$. With this in mind, we define its Choi state.

\begin{definition}
    The Choi state $\Upsilon^{\Tn}\in\Bcal(\Hcal_{\rmi_0}\otimes\Hcal_{\rmo_1}\otimes\cdots\otimes\Hcal_{\rmi_{n-1}}\otimes\Hcal_{\rmo_n})$ of the process tensor $\Tn$ is defined as
    \begin{equation}
        \Upsilon^{\Tn}\Def \qty(\bigotimes_{j=1}^{n}\Ical_{\rmi_{j-1}}\otimes \Tn)\qty(\bigotimes_{j=1}^{n}\Phi_{\rmi_{j-1},\rmi_{j-1}}).
    \end{equation}
\end{definition}

This definition is a straightforward generalization of Eq. \eqref{eq:choi_tt}. Also, the circuit generating this state is an extension of that shown in Fig. \ref{fig:choi_two}. Equations \eqref{eq:order1} and \eqref{eq:order2} are generalized as follows.

\begin{remark}
    In terms of the Choi state, the trace-preservation and time ordering properties of the process are given by the hierarchy of trace conditions
    \begin{equation}
    \label{eq:trace_conds}
        \tr_{\rmo_j}\qty[\Upsilon_{1:j}^{\Tt}] = \Upsilon_{1:j-1}^{\Tt}\otimes \Id_{\rmi_{j-1}}
    \end{equation}
    for all $1\le j\le n$, where
    \begin{equation}
    \label{eq:def_trace}
        \Upsilon_{1:j}^{\Tt}\Def\tr_{j+1:n}\qty[\Upsilon^{\Tt}]
    \end{equation}
    and $\tr_{j+1:n}$ is the trace over the subspaces $\{\rmi_{j},\rmo_{j+1},\cdots,\rmi_{n-1},\rmo_{n}\}$.
\end{remark}

Now, we move to the definition of the correlation quantifiers in the $n$-step scenario.

\begin{definition}
    The total correlations $\rmI$, Markovian correlations in the $j$-th step $\rmM_j$, total Markovian correlations $\rmM$, and non-Markovian correlations $\rmN$ of the $n$-step process tensor $\Tn$ are, respectively, defined as
    \begin{align}
        &\rmI(\Tn)\Def \rmI(\rmi_0:\rmo_1:\cdots:\rmi_{n-1}:\rmo_n)_{\Upsilon^{\Tn}},\\
        &\rmM_j(\Tn)\Def \rmI(\rmi_{j-1}:\rmo_j)_{\Upsilon^{\Tn}},\\
        &\rmM(\Tn)\Def \sum_{j=1}^{n}\rmM_j,\\
        &\rmN(\Tn)\Def \rmI(\rmi_0\rmo_1:\rmi_1\rmo_2:\cdots:\rmi_{n-1}\rmo_n)_{\Upsilon^{\Tn}}.
    \end{align}
\end{definition}

These definitions are equivalent to those of Ref. \cite{berk2023extracting}, which can be seen by writing the mutual information as the relative entropy between the state and the tensor product of its marginals, as shown in Eq. \eqref{eq:mutual_relative}.

Alternatively, the above definitions may be written explicitly in terms of the von Neumann entropies of the Choi state's marginals, which is done in the following remark.

\begin{remark}
    Let $\tr_{\overline{\rmA}}\qty[\Upsilon]$ denote the partial trace over all subsystems of $\Upsilon$ besides $\rmA$ and define
    \begin{align}
        &S_{\rmi_j}\Def S\qty(\tr_{\overline{\rmi_j}}\qty[\Upsilon^{\Tn}]),\\
        &S_{\rmo_j}\Def S\qty(\tr_{\overline{\rmo_j}}\qty[\Upsilon^{\Tn}]),\\
        &S_{j}\Def S\qty(\tr_{\overline{\rmi_{j-1}\rmo_j}}\qty[\Upsilon^{\Tn}]),\\
        &S_{\overline{j}}\Def S\qty(\tr_{\rmi_{k-1}\rmo_j}\qty[\Upsilon^{\Tn}])
    \end{align}
    and recalling Eq. \eqref{eq:def_trace},
    \begin{equation}
        S_{1:j}\Def S\qty(\Upsilon^{\Tn}_{1:j}).
    \end{equation}
Then, the correlation quantifiers may be rewritten as   
    \begin{align}
        &\rmI= \sum_{j=1}^{n}\qty(S_{\rmi_{j-1}}+S_{\rmo_{j}})-S_{1:n},\\
        &\rmM_j= S_{\rmi_{j-1}}+S_{\rmo_{j}}-S_{j},\\
        &\rmM= \sum_{j=1}^{n}\qty(S_{\rmi_{j-1}}+S_{\rmo_j}-S_{j}),\\
        &\rmN= \sum_{j=1}^{n}S_{j}-S_{1:n}.
    \end{align}
\end{remark}

Importantly, by comparing the above equations we readily obtain the additive property
    \begin{equation}
        \rmI=\rmM+\rmN.
    \end{equation}
Finally, we also define the complement of $\rmM_j$,
    \begin{equation}
        \overline{\rmM_j}\Def 2\ln d-\rmM_j.
    \end{equation}

Having set the stage with the definitions, we now proceed to the results.

\begin{proposition}
\label{prop:incomplete}
    For any (not necessarily time-ordered) process matrix $\Tn$, the following set of conditions holds,
    \begin{equation}
        \rmN\le 2\sum_{j\ne k}\overline{\rmM_j},
    \end{equation}
    for all $1\le k\le n$.
\end{proposition}
The proof can be found in Sec. \ref{app:prop1}.

\begin{example}
    For a four-step process, the conditions of Proposition \ref{prop:incomplete} are explicitly given by
    \begin{equation}
    \rmN\le
    \begin{cases}
        &2\qty(\overline{\rmM_2}+\overline{\rmM_3}+\overline{\rmM_4}),\\
        &2\qty(\overline{\rmM_1}+\overline{\rmM_3}+\overline{\rmM_4}),\\
        &2\qty(\overline{\rmM_1}+\overline{\rmM_2}+\overline{\rmM_4}),\\
        &2\qty(\overline{\rmM_1}+\overline{\rmM_2}+\overline{\rmM_3}).
    \end{cases}
\end{equation}
\end{example}

Nevertheless, this set of conditions may be refined if we take the time ordering of the process into account. That is done in the following proposition.

\begin{proposition}
\label{prop:complete}
    For any $n$-step process tensor $\Tn$, the following set of conditions holds:
    \begin{equation}
        \rmN\le 2\sum_{j<k}\overline{\rmM_j} + \sum_{j>k}\overline{\rmM_j}
    \end{equation}
    for all $1\le k\le n$.
\end{proposition}
The proof can be found in Sec \ref{app:prop2}.

\begin{example}
    For a four-step process, the conditions of Proposition \ref{prop:complete} are explicitly given by
    \begin{equation}
    \rmN\le
    \begin{cases}
        &\overline{\rmM_2}+\overline{\rmM_3}+\overline{\rmM_4},\\
        &2\overline{\rmM_1}+\overline{\rmM_3}+\overline{\rmM_4},\\
        &2\overline{\rmM_1}+2\overline{\rmM_2}+\overline{\rmM_4},\\
        &2\overline{\rmM_1}+2\overline{\rmM_2}+2\overline{\rmM_3}.
    \end{cases}
\end{equation}
\end{example}

Having proved the set of conditions of Proposition \ref{prop:complete}, we now derive two theorems concerning the non-Markovianity of general multitime quantum processes.

\begin{theorem}
    The maximum non-Markovianity $\rmN$ of any $n$-step process tensor $\Tn$ is $2(n-1)\ln d$.
\end{theorem}
The proof can be found in Sec \ref{app:thm1}.

\begin{example}
\label{ex:n_swap}
    Consider the $n$-step process associated with the Choi state
    \begin{equation}
        \Upsilon^{\Tn}=\Id_{\rmo_i}\otimes\Phi_{\rmi_0\rmo_2}\otimes\Phi_{\rmi_1\rmo_3}\otimes\cdots\otimes\Phi_{\rmi_{n-2}\rmo_{n}}\otimes\Id_{\rmi_{n-1}}.
    \end{equation}
    This is a generalization to $n$-steps of the process in Fig. \ref{fig:nm_dep} with $p=1$, such that the output of the $(j+1)$-th step is equal to the input of the $j$-th step. The first output is $\Id$, and the last input is discarded. It is straightforward to show that for this process
    \begin{equation}
        \rmN=2(n-1)\ln d;
    \end{equation}
    that is, it is maximally non-Markovian. Also, we have that $\rmM=0$, such that $\rmI=\rmN$.
\end{example}

Now, we derive how the non-Markovianity upper bounds the maximum possible Markovianity of a given process.

\begin{manualtheorem}{2}\label{thm:M_N}
For any $n$-step process tensor $\Tn$ it holds that
\begin{equation}
    \rmM\le 2n\ln d - \frac{2^n-1}{2^n-2}\rmN.
\end{equation}
\end{manualtheorem}
The proof can be found in Sec \ref{app:thm2}.

Alternatively, we may rewrite this result showing how the non-Markovianity limits the total correlations of the process.

\begin{manualtheorem}{2$^\prime$}
For any $n$-step process tensor $\Tn$ it holds that
\begin{equation}
    \rmI\le 2n\ln d - \frac{1}{2^n-2}\rmN.
\end{equation}
\end{manualtheorem}
The proof can be found in Sec \ref{app:thm2'}.

This last theorem shows that a non-Markovian process cannot have the maximum amount of total correlations. However, the price paid in total correlations to have a given amount of non-Markovianity decays exponentially with the number of steps of the process. This implies that a highly non-Markovian process with many steps might achieve an amount of total correlations relatively close to that of a unitary process with a similar number of steps.

\section{Conclusions}\label{sec:conclusion}

We have shown how to establish useful bounds relating the different types of temporal correlations present in general multitime quantum processes. We used the notion of information exchange between the system and the environment and the time ordering of the process to prove a set of inequalities relating Markovian correlations in individual steps to the global non-Markovianity. From these, we derived relevant results concerning the correlation quantifiers, like the maximum possible value of the non-Markovianity and how it limits the total correlations of the process depending on the number of steps.

These results could be applied, for example, to bound the efficiency of protocols like those in Ref. \cite{berk2023extracting}. Since the goal of their technique is to maximize the input-output correlation of the coarse-grained version of the process, which is upper bounded by the total correlations of the original process, one could use our Theorem $2^\prime$ to show how the non-Markovianity of the process limits the performance of the optimized dynamical decoupling protocol described there. Furthermore, our results could be used to provide fundamental bounds to the amount of memory resources one must expend to simulate a given quantum process, independent of the specific experimental setup employed, similar to what Ref. \cite{faist2019thermodynamic} does with thermodynamic resources.

Therefore, given the importance of dealing with temporal correlations in quantum processes, we hope these results help us pave the way towards a deeper understanding of informational flow in quantum systems, perhaps leading to significant improvements of state-of-the-art quantum information processing technologies.

\begin{acknowledgments}
G.Z. acknowledges financial support from FAPESP (Grant No. 2022/00993-9), and D.O.S.-P. acknowledges the support from the funding agencies CNPq (Grant No. 304891/2022-3), FAPESP (Grant No. 2017/03727-0), and the Brazilian National Institute of Science and Technology of Quantum Information (INCT-IQ) Grant No. 465469/2014-0.
\end{acknowledgments}

\appendix

\section{Proofs}

\subsection{Proposition 1}\label{app:prop1}

\begin{manualproposition}{1}
    For any (not necessarily time-ordered) process matrix $\Tn$, the following set of conditions holds:
    \begin{equation}
        \rmN\le 2\sum_{j\ne k}\overline{\rmM_j}
    \end{equation}
    for all $1\le k\le n$.
\end{manualproposition}
\begin{proof}
    From Eq. \eqref{eq:AL} we have
    \begin{equation}
        |S_{\overline{k}}-S_k|\le S_{1:n},
    \end{equation}
    which implies
    \begin{equation}
        S_k-S_{1:n}\le S_{\overline{k}}.
    \end{equation}
    Also, the subadditivity of von Neumann entropy $S_\rmAB\le S_\rmA+S_\rmB$ \cite{nielsen2010quantum,wilde2013quantum} yields
    \begin{equation}
        S_{\overline{k}}\le \sum_{j\ne k}S_j.
    \end{equation}
    Applying this to the non-Markovianity, we obtain, for any $1\le k\le n$,
    \begin{align}
        \rmN&=\sum_{j=1}^{n}S_{j}-S_{1:n}\\
        &\le \sum_{j\ne k}S_{j} +S_{\overline{k}}\\
        &= 2\sum_{j\ne k}S_j\\
        &\le 2\sum_{j\ne k}\overline{\rmM_j}.
    \end{align}
\end{proof}

\subsection{Proposition 2}\label{app:prop2}

\begin{manualproposition}{2}
    For any $n$-step process tensor $\Tn$, the following set of conditions holds;
    \begin{equation}
        \rmN\le 2\sum_{j<k}\overline{\rmM_j} + \sum_{j>k}\overline{\rmM_j}
    \end{equation}
    for all $1\le k\le n$.
\end{manualproposition}
\begin{proof}
    The trace conditions of Eq. \eqref{eq:trace_conds} imply
    \begin{equation}
        S_{1:k-1,\rmi_{k-1}} = S_{1:k-1}+\ln d,
    \end{equation}
    so we have
    \begin{align}
        S_{1:k-1} &= S_{1:k-1,\rmi_{k-1}} -\ln d\\
        &\le S_{1:k}+S_{\rmo_{k}} -\ln d\\
        &\le S_{1:k}
    \end{align}
    since $S_{1:k-1,\rmi_{k-1}}-S_{\rmo_{k}}\le S_{1:k}$ holds from Eq. \eqref{eq:AL} and also $S_{\rmo_{k}}\le\ln d$. Then, we can write
    \begin{align}
        S_k-S_{1:n}&\le S_k-S_{1:k}\\
        &\le S_{1:k-1}\\
        &\le \sum_{j<k}S_j,
    \end{align}
    using $S_k-S_{1:k-1}\le S_{1:k}$, again from Eq. \eqref{eq:AL}, and subadditivity. Finally, for the non-Markovianity
    \begin{align}
        \rmN&=\sum_{j=1}^{n}S_{j}-S_{1:n}\\
        &= \sum_{j\ne k}S_{j} + \qty(S_{k}-S_{1:n})\\
        &\le \sum_{j\ne k}S_{j} +\sum_{j<k}S_j\\
        &= 2\sum_{j<k}S_j+\sum_{j>k}S_j\\
        &\le 2\sum_{j<k}\overline{\rmM_j} + \sum_{j>k}\overline{\rmM_j}.
    \end{align}
\end{proof}

\subsection{Theorem 1}\label{app:thm1}

\begin{manualtheorem}{1}
    The maximum non-Markovianity $\rmN$ of any $n$-step process tensor $\Tn$ is $2(n-1)\ln d$.
\end{manualtheorem}
\begin{proof}
    From the set of conditions in Proposition \ref{prop:complete}, choose $k=1$, yielding
    \begin{align}
        \rmN&\le \sum_{j=2}^{n}\overline{\rmM_j}\\
        &= 2(n-1)\ln d - \sum_{j=2}^{n}\rmM_j\\
        &\le 2(n-1)\ln d.
    \end{align}
\end{proof}

\subsection{Theorem 2}\label{app:thm2}

\begin{manualtheorem}{2}
For any $n$-step process tensor $\Tn$ it holds that
\begin{equation}
    \rmM\le 2n\ln d - \frac{2^n-1}{2^n-2}\rmN.
\end{equation}
\end{manualtheorem}
\begin{proof}
    Take each condition from Proposition \ref{prop:complete}, multiply both sides by $2^{n-k}$, and sum for every $1\le k\le n$, yielding
    \begin{equation}
        \sum_{k=1}^n 2^{n-k} \rmN \le \sum_{k=1}^n 2^{n-k} \qty[2\sum_{j=1}^{k-1}\overline{\rmM_j}+ \sum_{j=k+1}^n\overline{\rmM_j}];
    \end{equation}
    after some tedious calculations we obtain
    \begin{align}
        \qty(2^n-1)\rmN&\le \qty(2^n-2)\sum_{j=1}^{n}\overline{\rmM_j}\\
        &= \qty(2^n-2)\qty(2n\ln d - \rmM),
    \end{align}
    which, after isolating $\rmM$, results in the inequality of the theorem.
\end{proof}

\subsection{Theorem 2$^\prime$}\label{app:thm2'}

\begin{manualtheorem}{2$^\prime$}
For any $n$-step process tensor $\Tn$ it holds that
\begin{equation}
    \rmI\le 2n\ln d - \frac{1}{2^n-2}\rmN.
\end{equation}
\end{manualtheorem}
\begin{proof}
    Take Theorem \ref{thm:M_N} and add $\rmN$ to both sides. On the left-hand side we use $\rmM+\rmN=\rmI$, obtaining the inequality of the theorem.
\end{proof}


\begin{thebibliography}{72}%
\makeatletter
\providecommand \@ifxundefined [1]{%
 \@ifx{#1\undefined}
}%
\providecommand \@ifnum [1]{%
 \ifnum #1\expandafter \@firstoftwo
 \else \expandafter \@secondoftwo
 \fi
}%
\providecommand \@ifx [1]{%
 \ifx #1\expandafter \@firstoftwo
 \else \expandafter \@secondoftwo
 \fi
}%
\providecommand \natexlab [1]{#1}%
\providecommand \enquote  [1]{``#1''}%
\providecommand \bibnamefont  [1]{#1}%
\providecommand \bibfnamefont [1]{#1}%
\providecommand \citenamefont [1]{#1}%
\providecommand \href@noop [0]{\@secondoftwo}%
\providecommand \href [0]{\begingroup \@sanitize@url \@href}%
\providecommand \@href[1]{\@@startlink{#1}\@@href}%
\providecommand \@@href[1]{\endgroup#1\@@endlink}%
\providecommand \@sanitize@url [0]{\catcode `\\12\catcode `\$12\catcode `\&12\catcode `\#12\catcode `\^12\catcode `\_12\catcode `\%12\relax}%
\providecommand \@@startlink[1]{}%
\providecommand \@@endlink[0]{}%
\providecommand \url  [0]{\begingroup\@sanitize@url \@url }%
\providecommand \@url [1]{\endgroup\@href {#1}{\urlprefix }}%
\providecommand \urlprefix  [0]{URL }%
\providecommand \Eprint [0]{\href }%
\providecommand \doibase [0]{https://doi.org/}%
\providecommand \selectlanguage [0]{\@gobble}%
\providecommand \bibinfo  [0]{\@secondoftwo}%
\providecommand \bibfield  [0]{\@secondoftwo}%
\providecommand \translation [1]{[#1]}%
\providecommand \BibitemOpen [0]{}%
\providecommand \bibitemStop [0]{}%
\providecommand \bibitemNoStop [0]{.\EOS\space}%
\providecommand \EOS [0]{\spacefactor3000\relax}%
\providecommand \BibitemShut  [1]{\csname bibitem#1\endcsname}%
\let\auto@bib@innerbib\@empty
%</preamble>
\bibitem [{\citenamefont {Breuer}\ and\ \citenamefont {Petruccione}(2002)}]{breuer2002theory}%
  \BibitemOpen
  \bibfield  {author} {\bibinfo {author} {\bibfnamefont {H.-P.}\ \bibnamefont {Breuer}}\ and\ \bibinfo {author} {\bibfnamefont {F.}~\bibnamefont {Petruccione}},\ }\href@noop {} {\emph {\bibinfo {title} {The theory of open quantum systems}}}\ (\bibinfo  {publisher} {Oxford University Press, USA},\ \bibinfo {year} {2002})\BibitemShut {NoStop}%
\bibitem [{\citenamefont {Rivas}\ and\ \citenamefont {Huelga}(2012)}]{rivas2012open}%
  \BibitemOpen
  \bibfield  {author} {\bibinfo {author} {\bibfnamefont {A.}~\bibnamefont {Rivas}}\ and\ \bibinfo {author} {\bibfnamefont {S.~F.}\ \bibnamefont {Huelga}},\ }\href@noop {} {\emph {\bibinfo {title} {Open quantum systems}}},\ Vol.~\bibinfo {volume} {10}\ (\bibinfo  {publisher} {Springer},\ \bibinfo {year} {2012})\BibitemShut {NoStop}%
\bibitem [{\citenamefont {Nielsen}\ and\ \citenamefont {Chuang}(2010)}]{nielsen2010quantum}%
  \BibitemOpen
  \bibfield  {author} {\bibinfo {author} {\bibfnamefont {M.~A.}\ \bibnamefont {Nielsen}}\ and\ \bibinfo {author} {\bibfnamefont {I.~L.}\ \bibnamefont {Chuang}},\ }\href@noop {} {\emph {\bibinfo {title} {Quantum computation and quantum information}}}\ (\bibinfo  {publisher} {Cambridge university press},\ \bibinfo {year} {2010})\BibitemShut {NoStop}%
\bibitem [{\citenamefont {Wilde}(2013)}]{wilde2013quantum}%
  \BibitemOpen
  \bibfield  {author} {\bibinfo {author} {\bibfnamefont {M.~M.}\ \bibnamefont {Wilde}},\ }\href@noop {} {\emph {\bibinfo {title} {Quantum information theory}}}\ (\bibinfo  {publisher} {Cambridge university press},\ \bibinfo {year} {2013})\BibitemShut {NoStop}%
\bibitem [{\citenamefont {Preskill}(2018)}]{preskill2018quantum}%
  \BibitemOpen
  \bibfield  {author} {\bibinfo {author} {\bibfnamefont {J.}~\bibnamefont {Preskill}},\ }\href {https://doi.org/10.22331/q-2018-08-06-79} {\bibfield  {journal} {\bibinfo  {journal} {{Quantum}}\ }\textbf {\bibinfo {volume} {2}},\ \bibinfo {pages} {79} (\bibinfo {year} {2018})}\BibitemShut {NoStop}%
\bibitem [{\citenamefont {Rivas}\ \emph {et~al.}(2014)\citenamefont {Rivas}, \citenamefont {Huelga},\ and\ \citenamefont {Plenio}}]{rivas2014quantum}%
  \BibitemOpen
  \bibfield  {author} {\bibinfo {author} {\bibfnamefont {{\'A}.}~\bibnamefont {Rivas}}, \bibinfo {author} {\bibfnamefont {S.~F.}\ \bibnamefont {Huelga}},\ and\ \bibinfo {author} {\bibfnamefont {M.~B.}\ \bibnamefont {Plenio}},\ }\href {https://doi.org/10.1088/0034-4885/77/9/094001} {\bibfield  {journal} {\bibinfo  {journal} {Reports on Progress in Physics}\ }\textbf {\bibinfo {volume} {77}},\ \bibinfo {pages} {094001} (\bibinfo {year} {2014})}\BibitemShut {NoStop}%
\bibitem [{\citenamefont {Breuer}\ \emph {et~al.}(2016)\citenamefont {Breuer}, \citenamefont {Laine}, \citenamefont {Piilo},\ and\ \citenamefont {Vacchini}}]{breuer2016colloquium}%
  \BibitemOpen
  \bibfield  {author} {\bibinfo {author} {\bibfnamefont {H.-P.}\ \bibnamefont {Breuer}}, \bibinfo {author} {\bibfnamefont {E.-M.}\ \bibnamefont {Laine}}, \bibinfo {author} {\bibfnamefont {J.}~\bibnamefont {Piilo}},\ and\ \bibinfo {author} {\bibfnamefont {B.}~\bibnamefont {Vacchini}},\ }\href {https://doi.org/10.1103/RevModPhys.88.021002} {\bibfield  {journal} {\bibinfo  {journal} {Rev. Mod. Phys.}\ }\textbf {\bibinfo {volume} {88}},\ \bibinfo {pages} {021002} (\bibinfo {year} {2016})}\BibitemShut {NoStop}%
\bibitem [{\citenamefont {de~Vega}\ and\ \citenamefont {Alonso}(2017)}]{vega2017dynamics}%
  \BibitemOpen
  \bibfield  {author} {\bibinfo {author} {\bibfnamefont {I.}~\bibnamefont {de~Vega}}\ and\ \bibinfo {author} {\bibfnamefont {D.}~\bibnamefont {Alonso}},\ }\href {https://doi.org/10.1103/RevModPhys.89.015001} {\bibfield  {journal} {\bibinfo  {journal} {Rev. Mod. Phys.}\ }\textbf {\bibinfo {volume} {89}},\ \bibinfo {pages} {015001} (\bibinfo {year} {2017})}\BibitemShut {NoStop}%
\bibitem [{\citenamefont {Li}\ \emph {et~al.}(2018)\citenamefont {Li}, \citenamefont {Hall},\ and\ \citenamefont {Wiseman}}]{li2018concepts}%
  \BibitemOpen
  \bibfield  {author} {\bibinfo {author} {\bibfnamefont {L.}~\bibnamefont {Li}}, \bibinfo {author} {\bibfnamefont {M.~J.}\ \bibnamefont {Hall}},\ and\ \bibinfo {author} {\bibfnamefont {H.~M.}\ \bibnamefont {Wiseman}},\ }\href {https://doi.org/10.1016/j.physrep.2018.07.001} {\bibfield  {journal} {\bibinfo  {journal} {Physics Reports}\ }\textbf {\bibinfo {volume} {759}},\ \bibinfo {pages} {1} (\bibinfo {year} {2018})}\BibitemShut {NoStop}%
\bibitem [{\citenamefont {Pollock}\ \emph {et~al.}(2018{\natexlab{a}})\citenamefont {Pollock}, \citenamefont {Rodr{\'\i}guez-Rosario}, \citenamefont {Frauenheim}, \citenamefont {Paternostro},\ and\ \citenamefont {Modi}}]{pollock2018non}%
  \BibitemOpen
  \bibfield  {author} {\bibinfo {author} {\bibfnamefont {F.~A.}\ \bibnamefont {Pollock}}, \bibinfo {author} {\bibfnamefont {C.}~\bibnamefont {Rodr{\'\i}guez-Rosario}}, \bibinfo {author} {\bibfnamefont {T.}~\bibnamefont {Frauenheim}}, \bibinfo {author} {\bibfnamefont {M.}~\bibnamefont {Paternostro}},\ and\ \bibinfo {author} {\bibfnamefont {K.}~\bibnamefont {Modi}},\ }\href {https://doi.org/10.1103/PhysRevA.97.012127} {\bibfield  {journal} {\bibinfo  {journal} {Physical Review A}\ }\textbf {\bibinfo {volume} {97}},\ \bibinfo {pages} {012127} (\bibinfo {year} {2018}{\natexlab{a}})}\BibitemShut {NoStop}%
\bibitem [{\citenamefont {Milz}\ \emph {et~al.}(2020{\natexlab{a}})\citenamefont {Milz}, \citenamefont {Sakuldee}, \citenamefont {Pollock},\ and\ \citenamefont {Modi}}]{milz2020kolmogorov}%
  \BibitemOpen
  \bibfield  {author} {\bibinfo {author} {\bibfnamefont {S.}~\bibnamefont {Milz}}, \bibinfo {author} {\bibfnamefont {F.}~\bibnamefont {Sakuldee}}, \bibinfo {author} {\bibfnamefont {F.~A.}\ \bibnamefont {Pollock}},\ and\ \bibinfo {author} {\bibfnamefont {K.}~\bibnamefont {Modi}},\ }\href {https://doi.org/10.22331/q-2020-04-20-255} {\bibfield  {journal} {\bibinfo  {journal} {Quantum}\ }\textbf {\bibinfo {volume} {4}},\ \bibinfo {pages} {255} (\bibinfo {year} {2020}{\natexlab{a}})}\BibitemShut {NoStop}%
\bibitem [{\citenamefont {Pollock}\ \emph {et~al.}(2018{\natexlab{b}})\citenamefont {Pollock}, \citenamefont {Rodr{\'\i}guez-Rosario}, \citenamefont {Frauenheim}, \citenamefont {Paternostro},\ and\ \citenamefont {Modi}}]{pollock2018operational}%
  \BibitemOpen
  \bibfield  {author} {\bibinfo {author} {\bibfnamefont {F.~A.}\ \bibnamefont {Pollock}}, \bibinfo {author} {\bibfnamefont {C.}~\bibnamefont {Rodr{\'\i}guez-Rosario}}, \bibinfo {author} {\bibfnamefont {T.}~\bibnamefont {Frauenheim}}, \bibinfo {author} {\bibfnamefont {M.}~\bibnamefont {Paternostro}},\ and\ \bibinfo {author} {\bibfnamefont {K.}~\bibnamefont {Modi}},\ }\href {https://doi.org/10.1103/PhysRevLett.120.040405} {\bibfield  {journal} {\bibinfo  {journal} {Physical review letters}\ }\textbf {\bibinfo {volume} {120}},\ \bibinfo {pages} {040405} (\bibinfo {year} {2018}{\natexlab{b}})}\BibitemShut {NoStop}%
\bibitem [{\citenamefont {Milz}\ \emph {et~al.}(2019)\citenamefont {Milz}, \citenamefont {Kim}, \citenamefont {Pollock},\ and\ \citenamefont {Modi}}]{milz2019completely}%
  \BibitemOpen
  \bibfield  {author} {\bibinfo {author} {\bibfnamefont {S.}~\bibnamefont {Milz}}, \bibinfo {author} {\bibfnamefont {M.}~\bibnamefont {Kim}}, \bibinfo {author} {\bibfnamefont {F.~A.}\ \bibnamefont {Pollock}},\ and\ \bibinfo {author} {\bibfnamefont {K.}~\bibnamefont {Modi}},\ }\href {https://doi.org/10.1103/PhysRevLett.123.040401} {\bibfield  {journal} {\bibinfo  {journal} {Physical review letters}\ }\textbf {\bibinfo {volume} {123}},\ \bibinfo {pages} {040401} (\bibinfo {year} {2019})}\BibitemShut {NoStop}%
\bibitem [{\citenamefont {Taranto}\ \emph {et~al.}(2019{\natexlab{a}})\citenamefont {Taranto}, \citenamefont {Pollock}, \citenamefont {Milz}, \citenamefont {Tomamichel},\ and\ \citenamefont {Modi}}]{taranto2019quantum}%
  \BibitemOpen
  \bibfield  {author} {\bibinfo {author} {\bibfnamefont {P.}~\bibnamefont {Taranto}}, \bibinfo {author} {\bibfnamefont {F.~A.}\ \bibnamefont {Pollock}}, \bibinfo {author} {\bibfnamefont {S.}~\bibnamefont {Milz}}, \bibinfo {author} {\bibfnamefont {M.}~\bibnamefont {Tomamichel}},\ and\ \bibinfo {author} {\bibfnamefont {K.}~\bibnamefont {Modi}},\ }\href {https://doi.org/10.1103/PhysRevLett.122.140401} {\bibfield  {journal} {\bibinfo  {journal} {Physical Review Letters}\ }\textbf {\bibinfo {volume} {122}},\ \bibinfo {pages} {140401} (\bibinfo {year} {2019}{\natexlab{a}})}\BibitemShut {NoStop}%
\bibitem [{\citenamefont {Taranto}\ \emph {et~al.}(2019{\natexlab{b}})\citenamefont {Taranto}, \citenamefont {Milz}, \citenamefont {Pollock},\ and\ \citenamefont {Modi}}]{taranto2019structure}%
  \BibitemOpen
  \bibfield  {author} {\bibinfo {author} {\bibfnamefont {P.}~\bibnamefont {Taranto}}, \bibinfo {author} {\bibfnamefont {S.}~\bibnamefont {Milz}}, \bibinfo {author} {\bibfnamefont {F.~A.}\ \bibnamefont {Pollock}},\ and\ \bibinfo {author} {\bibfnamefont {K.}~\bibnamefont {Modi}},\ }\href {https://doi.org/10.1103/PhysRevA.99.042108} {\bibfield  {journal} {\bibinfo  {journal} {Physical Review A}\ }\textbf {\bibinfo {volume} {99}},\ \bibinfo {pages} {042108} (\bibinfo {year} {2019}{\natexlab{b}})}\BibitemShut {NoStop}%
\bibitem [{\citenamefont {Figueroa-Romero}\ \emph {et~al.}(2019)\citenamefont {Figueroa-Romero}, \citenamefont {Modi},\ and\ \citenamefont {Pollock}}]{figueroa2019almost}%
  \BibitemOpen
  \bibfield  {author} {\bibinfo {author} {\bibfnamefont {P.}~\bibnamefont {Figueroa-Romero}}, \bibinfo {author} {\bibfnamefont {K.}~\bibnamefont {Modi}},\ and\ \bibinfo {author} {\bibfnamefont {F.~A.}\ \bibnamefont {Pollock}},\ }\href {https://doi.org/10.22331/q-2019-04-30-136} {\bibfield  {journal} {\bibinfo  {journal} {Quantum}\ }\textbf {\bibinfo {volume} {3}},\ \bibinfo {pages} {136} (\bibinfo {year} {2019})}\BibitemShut {NoStop}%
\bibitem [{\citenamefont {Milz}\ \emph {et~al.}(2020{\natexlab{b}})\citenamefont {Milz}, \citenamefont {Egloff}, \citenamefont {Taranto}, \citenamefont {Theurer}, \citenamefont {Plenio}, \citenamefont {Smirne},\ and\ \citenamefont {Huelga}}]{milz2020when}%
  \BibitemOpen
  \bibfield  {author} {\bibinfo {author} {\bibfnamefont {S.}~\bibnamefont {Milz}}, \bibinfo {author} {\bibfnamefont {D.}~\bibnamefont {Egloff}}, \bibinfo {author} {\bibfnamefont {P.}~\bibnamefont {Taranto}}, \bibinfo {author} {\bibfnamefont {T.}~\bibnamefont {Theurer}}, \bibinfo {author} {\bibfnamefont {M.~B.}\ \bibnamefont {Plenio}}, \bibinfo {author} {\bibfnamefont {A.}~\bibnamefont {Smirne}},\ and\ \bibinfo {author} {\bibfnamefont {S.~F.}\ \bibnamefont {Huelga}},\ }\href {https://doi.org/10.1103/PhysRevX.10.041049} {\bibfield  {journal} {\bibinfo  {journal} {Physical Review X}\ }\textbf {\bibinfo {volume} {10}},\ \bibinfo {pages} {041049} (\bibinfo {year} {2020}{\natexlab{b}})}\BibitemShut {NoStop}%
\bibitem [{\citenamefont {Milz}\ \emph {et~al.}(2021)\citenamefont {Milz}, \citenamefont {Spee}, \citenamefont {Xu}, \citenamefont {Pollock}, \citenamefont {Modi},\ and\ \citenamefont {G{\"u}hne}}]{milz2021genuine}%
  \BibitemOpen
  \bibfield  {author} {\bibinfo {author} {\bibfnamefont {S.}~\bibnamefont {Milz}}, \bibinfo {author} {\bibfnamefont {C.}~\bibnamefont {Spee}}, \bibinfo {author} {\bibfnamefont {Z.-P.}\ \bibnamefont {Xu}}, \bibinfo {author} {\bibfnamefont {F.~A.}\ \bibnamefont {Pollock}}, \bibinfo {author} {\bibfnamefont {K.}~\bibnamefont {Modi}},\ and\ \bibinfo {author} {\bibfnamefont {O.}~\bibnamefont {G{\"u}hne}},\ }\href {https://doi.org/10.21468/SciPostPhys.10.6.141} {\bibfield  {journal} {\bibinfo  {journal} {SciPost Physics}\ }\textbf {\bibinfo {volume} {10}},\ \bibinfo {pages} {141} (\bibinfo {year} {2021})}\BibitemShut {NoStop}%
\bibitem [{\citenamefont {Taranto}\ \emph {et~al.}(2021)\citenamefont {Taranto}, \citenamefont {Pollock},\ and\ \citenamefont {Modi}}]{taranto2021non}%
  \BibitemOpen
  \bibfield  {author} {\bibinfo {author} {\bibfnamefont {P.}~\bibnamefont {Taranto}}, \bibinfo {author} {\bibfnamefont {F.~A.}\ \bibnamefont {Pollock}},\ and\ \bibinfo {author} {\bibfnamefont {K.}~\bibnamefont {Modi}},\ }\href {https://doi.org/10.1038/s41534-021-00481-4} {\bibfield  {journal} {\bibinfo  {journal} {npj Quantum Information}\ }\textbf {\bibinfo {volume} {7}},\ \bibinfo {pages} {149} (\bibinfo {year} {2021})}\BibitemShut {NoStop}%
\bibitem [{\citenamefont {Figueroa-Romero}\ \emph {et~al.}(2021{\natexlab{a}})\citenamefont {Figueroa-Romero}, \citenamefont {Pollock},\ and\ \citenamefont {Modi}}]{figueroa2021markovianization}%
  \BibitemOpen
  \bibfield  {author} {\bibinfo {author} {\bibfnamefont {P.}~\bibnamefont {Figueroa-Romero}}, \bibinfo {author} {\bibfnamefont {F.~A.}\ \bibnamefont {Pollock}},\ and\ \bibinfo {author} {\bibfnamefont {K.}~\bibnamefont {Modi}},\ }\href {https://doi.org/10.1038/s42005-021-00629-w} {\bibfield  {journal} {\bibinfo  {journal} {Communications Physics}\ }\textbf {\bibinfo {volume} {4}},\ \bibinfo {pages} {127} (\bibinfo {year} {2021}{\natexlab{a}})}\BibitemShut {NoStop}%
\bibitem [{\citenamefont {Sakuldee}\ \emph {et~al.}(2022)\citenamefont {Sakuldee}, \citenamefont {Taranto},\ and\ \citenamefont {Milz}}]{sakuldee2022connecting}%
  \BibitemOpen
  \bibfield  {author} {\bibinfo {author} {\bibfnamefont {F.}~\bibnamefont {Sakuldee}}, \bibinfo {author} {\bibfnamefont {P.}~\bibnamefont {Taranto}},\ and\ \bibinfo {author} {\bibfnamefont {S.}~\bibnamefont {Milz}},\ }\href {https://doi.org/10.1103/PhysRevA.106.022416} {\bibfield  {journal} {\bibinfo  {journal} {Physical Review A}\ }\textbf {\bibinfo {volume} {106}},\ \bibinfo {pages} {022416} (\bibinfo {year} {2022})}\BibitemShut {NoStop}%
\bibitem [{\citenamefont {Capela}\ \emph {et~al.}(2022)\citenamefont {Capela}, \citenamefont {C{\'e}leri}, \citenamefont {Chaves},\ and\ \citenamefont {Modi}}]{capela2022quantum}%
  \BibitemOpen
  \bibfield  {author} {\bibinfo {author} {\bibfnamefont {M.}~\bibnamefont {Capela}}, \bibinfo {author} {\bibfnamefont {L.~C.}\ \bibnamefont {C{\'e}leri}}, \bibinfo {author} {\bibfnamefont {R.}~\bibnamefont {Chaves}},\ and\ \bibinfo {author} {\bibfnamefont {K.}~\bibnamefont {Modi}},\ }\href {https://doi.org/10.1103/PhysRevA.106.022218} {\bibfield  {journal} {\bibinfo  {journal} {Physical Review A}\ }\textbf {\bibinfo {volume} {106}},\ \bibinfo {pages} {022218} (\bibinfo {year} {2022})}\BibitemShut {NoStop}%
\bibitem [{\citenamefont {Taranto}\ \emph {et~al.}(2023{\natexlab{a}})\citenamefont {Taranto}, \citenamefont {Elliott},\ and\ \citenamefont {Milz}}]{taranto2023hidden}%
  \BibitemOpen
  \bibfield  {author} {\bibinfo {author} {\bibfnamefont {P.}~\bibnamefont {Taranto}}, \bibinfo {author} {\bibfnamefont {T.~J.}\ \bibnamefont {Elliott}},\ and\ \bibinfo {author} {\bibfnamefont {S.}~\bibnamefont {Milz}},\ }\href {https://doi.org/10.22331/q-2023-04-27-991} {\bibfield  {journal} {\bibinfo  {journal} {Quantum}\ }\textbf {\bibinfo {volume} {7}},\ \bibinfo {pages} {991} (\bibinfo {year} {2023}{\natexlab{a}})}\BibitemShut {NoStop}%
\bibitem [{\citenamefont {Taranto}\ \emph {et~al.}(2023{\natexlab{b}})\citenamefont {Taranto}, \citenamefont {Quintino}, \citenamefont {Murao},\ and\ \citenamefont {Milz}}]{taranto2023characterising}%
  \BibitemOpen
  \bibfield  {author} {\bibinfo {author} {\bibfnamefont {P.}~\bibnamefont {Taranto}}, \bibinfo {author} {\bibfnamefont {M.~T.}\ \bibnamefont {Quintino}}, \bibinfo {author} {\bibfnamefont {M.}~\bibnamefont {Murao}},\ and\ \bibinfo {author} {\bibfnamefont {S.}~\bibnamefont {Milz}},\ }\bibfield  {journal} {\bibinfo  {journal} {arXiv preprint arXiv:2307.11905}\ }\href {https://doi.org/10.48550/arXiv.2307.11905} {10.48550/arXiv.2307.11905} (\bibinfo {year} {2023}{\natexlab{b}})\BibitemShut {NoStop}%
\bibitem [{\citenamefont {van Kampen}(1992)}]{van1992stochastic}%
  \BibitemOpen
  \bibfield  {author} {\bibinfo {author} {\bibfnamefont {N.~G.}\ \bibnamefont {van Kampen}},\ }\href@noop {} {\emph {\bibinfo {title} {Stochastic processes in physics and chemistry}}},\ Vol.~\bibinfo {volume} {1}\ (\bibinfo  {publisher} {Elsevier},\ \bibinfo {year} {1992})\BibitemShut {NoStop}%
\bibitem [{\citenamefont {van Kampen}(1998)}]{van1998remarks}%
  \BibitemOpen
  \bibfield  {author} {\bibinfo {author} {\bibfnamefont {N.~G.}\ \bibnamefont {van Kampen}},\ }\href {https://doi.org/10.1590/S0103-97331998000200003} {\bibfield  {journal} {\bibinfo  {journal} {Brazilian Journal of Physics}\ }\textbf {\bibinfo {volume} {28}},\ \bibinfo {pages} {90} (\bibinfo {year} {1998})}\BibitemShut {NoStop}%
\bibitem [{\citenamefont {J\o{}rgensen}\ and\ \citenamefont {Pollock}(2019)}]{jorgensen2019exploiting}%
  \BibitemOpen
  \bibfield  {author} {\bibinfo {author} {\bibfnamefont {M.~R.}\ \bibnamefont {J\o{}rgensen}}\ and\ \bibinfo {author} {\bibfnamefont {F.~A.}\ \bibnamefont {Pollock}},\ }\href {https://doi.org/10.1103/PhysRevLett.123.240602} {\bibfield  {journal} {\bibinfo  {journal} {Phys. Rev. Lett.}\ }\textbf {\bibinfo {volume} {123}},\ \bibinfo {pages} {240602} (\bibinfo {year} {2019})}\BibitemShut {NoStop}%
\bibitem [{\citenamefont {J\o{}rgensen}\ and\ \citenamefont {Pollock}(2020)}]{jorgensen2020discrete}%
  \BibitemOpen
  \bibfield  {author} {\bibinfo {author} {\bibfnamefont {M.~R.}\ \bibnamefont {J\o{}rgensen}}\ and\ \bibinfo {author} {\bibfnamefont {F.~A.}\ \bibnamefont {Pollock}},\ }\href {https://doi.org/10.1103/PhysRevA.102.052206} {\bibfield  {journal} {\bibinfo  {journal} {Phys. Rev. A}\ }\textbf {\bibinfo {volume} {102}},\ \bibinfo {pages} {052206} (\bibinfo {year} {2020})}\BibitemShut {NoStop}%
\bibitem [{\citenamefont {Xiang}\ \emph {et~al.}(2021)\citenamefont {Xiang}, \citenamefont {Zong}, \citenamefont {Zhan}, \citenamefont {Fei}, \citenamefont {Run}, \citenamefont {Wu}, \citenamefont {Jin}, \citenamefont {Jia}, \citenamefont {Duan}, \citenamefont {Wu} \emph {et~al.}}]{xiang2021quantify}%
  \BibitemOpen
  \bibfield  {author} {\bibinfo {author} {\bibfnamefont {L.}~\bibnamefont {Xiang}}, \bibinfo {author} {\bibfnamefont {Z.}~\bibnamefont {Zong}}, \bibinfo {author} {\bibfnamefont {Z.}~\bibnamefont {Zhan}}, \bibinfo {author} {\bibfnamefont {Y.}~\bibnamefont {Fei}}, \bibinfo {author} {\bibfnamefont {C.}~\bibnamefont {Run}}, \bibinfo {author} {\bibfnamefont {Y.}~\bibnamefont {Wu}}, \bibinfo {author} {\bibfnamefont {W.}~\bibnamefont {Jin}}, \bibinfo {author} {\bibfnamefont {Z.}~\bibnamefont {Jia}}, \bibinfo {author} {\bibfnamefont {P.}~\bibnamefont {Duan}}, \bibinfo {author} {\bibfnamefont {J.}~\bibnamefont {Wu}}, \emph {et~al.},\ }\bibfield  {journal} {\bibinfo  {journal} {arXiv preprint arXiv:2105.03333}\ }\href {https://doi.org/10.48550/arXiv.2105.03333} {10.48550/arXiv.2105.03333} (\bibinfo {year} {2021})\BibitemShut {NoStop}%
\bibitem [{\citenamefont {Cygorek}\ \emph {et~al.}(2022)\citenamefont {Cygorek}, \citenamefont {Cosacchi}, \citenamefont {Vagov}, \citenamefont {Axt}, \citenamefont {Lovett}, \citenamefont {Keeling},\ and\ \citenamefont {Gauger}}]{cygorek2022simulation}%
  \BibitemOpen
  \bibfield  {author} {\bibinfo {author} {\bibfnamefont {M.}~\bibnamefont {Cygorek}}, \bibinfo {author} {\bibfnamefont {M.}~\bibnamefont {Cosacchi}}, \bibinfo {author} {\bibfnamefont {A.}~\bibnamefont {Vagov}}, \bibinfo {author} {\bibfnamefont {V.~M.}\ \bibnamefont {Axt}}, \bibinfo {author} {\bibfnamefont {B.~W.}\ \bibnamefont {Lovett}}, \bibinfo {author} {\bibfnamefont {J.}~\bibnamefont {Keeling}},\ and\ \bibinfo {author} {\bibfnamefont {E.~M.}\ \bibnamefont {Gauger}},\ }\href {https://doi.org/10.1038/s41567-022-01544-9} {\bibfield  {journal} {\bibinfo  {journal} {Nature Physics}\ }\textbf {\bibinfo {volume} {18}},\ \bibinfo {pages} {662} (\bibinfo {year} {2022})}\BibitemShut {NoStop}%
\bibitem [{\citenamefont {Fowler-Wright}\ \emph {et~al.}(2022)\citenamefont {Fowler-Wright}, \citenamefont {Lovett},\ and\ \citenamefont {Keeling}}]{fowler2022efficient}%
  \BibitemOpen
  \bibfield  {author} {\bibinfo {author} {\bibfnamefont {P.}~\bibnamefont {Fowler-Wright}}, \bibinfo {author} {\bibfnamefont {B.~W.}\ \bibnamefont {Lovett}},\ and\ \bibinfo {author} {\bibfnamefont {J.}~\bibnamefont {Keeling}},\ }\href {https://doi.org/10.1103/PhysRevLett.129.173001} {\bibfield  {journal} {\bibinfo  {journal} {Phys. Rev. Lett.}\ }\textbf {\bibinfo {volume} {129}},\ \bibinfo {pages} {173001} (\bibinfo {year} {2022})}\BibitemShut {NoStop}%
\bibitem [{\citenamefont {Gribben}\ \emph {et~al.}(2022)\citenamefont {Gribben}, \citenamefont {Strathearn}, \citenamefont {Fux}, \citenamefont {Kirton},\ and\ \citenamefont {Lovett}}]{gribben2022using}%
  \BibitemOpen
  \bibfield  {author} {\bibinfo {author} {\bibfnamefont {D.}~\bibnamefont {Gribben}}, \bibinfo {author} {\bibfnamefont {A.}~\bibnamefont {Strathearn}}, \bibinfo {author} {\bibfnamefont {G.~E.}\ \bibnamefont {Fux}}, \bibinfo {author} {\bibfnamefont {P.}~\bibnamefont {Kirton}},\ and\ \bibinfo {author} {\bibfnamefont {B.~W.}\ \bibnamefont {Lovett}},\ }\href {https://doi.org/10.22331/q-2022-10-25-847} {\bibfield  {journal} {\bibinfo  {journal} {{Quantum}}\ }\textbf {\bibinfo {volume} {6}},\ \bibinfo {pages} {847} (\bibinfo {year} {2022})}\BibitemShut {NoStop}%
\bibitem [{\citenamefont {Cygorek}\ \emph {et~al.}(2023)\citenamefont {Cygorek}, \citenamefont {Keeling}, \citenamefont {Lovett},\ and\ \citenamefont {Gauger}}]{cygorek2023sublinear}%
  \BibitemOpen
  \bibfield  {author} {\bibinfo {author} {\bibfnamefont {M.}~\bibnamefont {Cygorek}}, \bibinfo {author} {\bibfnamefont {J.}~\bibnamefont {Keeling}}, \bibinfo {author} {\bibfnamefont {B.~W.}\ \bibnamefont {Lovett}},\ and\ \bibinfo {author} {\bibfnamefont {E.~M.}\ \bibnamefont {Gauger}},\ }\bibfield  {journal} {\bibinfo  {journal} {arXiv preprint arXiv:2304.05291}\ }\href {https://doi.org/10.48550/arXiv.2304.05291} {10.48550/arXiv.2304.05291} (\bibinfo {year} {2023})\BibitemShut {NoStop}%
\bibitem [{\citenamefont {Fux}\ \emph {et~al.}(2023)\citenamefont {Fux}, \citenamefont {Kilda}, \citenamefont {Lovett},\ and\ \citenamefont {Keeling}}]{fux2023tensor}%
  \BibitemOpen
  \bibfield  {author} {\bibinfo {author} {\bibfnamefont {G.~E.}\ \bibnamefont {Fux}}, \bibinfo {author} {\bibfnamefont {D.}~\bibnamefont {Kilda}}, \bibinfo {author} {\bibfnamefont {B.~W.}\ \bibnamefont {Lovett}},\ and\ \bibinfo {author} {\bibfnamefont {J.}~\bibnamefont {Keeling}},\ }\href {https://doi.org/10.1103/PhysRevResearch.5.033078} {\bibfield  {journal} {\bibinfo  {journal} {Phys. Rev. Res.}\ }\textbf {\bibinfo {volume} {5}},\ \bibinfo {pages} {033078} (\bibinfo {year} {2023})}\BibitemShut {NoStop}%
\bibitem [{\citenamefont {Figueroa-Romero}\ \emph {et~al.}(2021{\natexlab{b}})\citenamefont {Figueroa-Romero}, \citenamefont {Modi}, \citenamefont {Harris}, \citenamefont {Stace},\ and\ \citenamefont {Hsieh}}]{figueroa2021randomized}%
  \BibitemOpen
  \bibfield  {author} {\bibinfo {author} {\bibfnamefont {P.}~\bibnamefont {Figueroa-Romero}}, \bibinfo {author} {\bibfnamefont {K.}~\bibnamefont {Modi}}, \bibinfo {author} {\bibfnamefont {R.~J.}\ \bibnamefont {Harris}}, \bibinfo {author} {\bibfnamefont {T.~M.}\ \bibnamefont {Stace}},\ and\ \bibinfo {author} {\bibfnamefont {M.-H.}\ \bibnamefont {Hsieh}},\ }\href {https://doi.org/10.1103/PRXQuantum.2.040351} {\bibfield  {journal} {\bibinfo  {journal} {PRX Quantum}\ }\textbf {\bibinfo {volume} {2}},\ \bibinfo {pages} {040351} (\bibinfo {year} {2021}{\natexlab{b}})}\BibitemShut {NoStop}%
\bibitem [{\citenamefont {Figueroa-Romero}\ \emph {et~al.}(2022)\citenamefont {Figueroa-Romero}, \citenamefont {Modi},\ and\ \citenamefont {Hsieh}}]{figueroa2022towards}%
  \BibitemOpen
  \bibfield  {author} {\bibinfo {author} {\bibfnamefont {P.}~\bibnamefont {Figueroa-Romero}}, \bibinfo {author} {\bibfnamefont {K.}~\bibnamefont {Modi}},\ and\ \bibinfo {author} {\bibfnamefont {M.-H.}\ \bibnamefont {Hsieh}},\ }\href {https://doi.org/10.22331/q-2022-12-01-868} {\bibfield  {journal} {\bibinfo  {journal} {Quantum}\ }\textbf {\bibinfo {volume} {6}},\ \bibinfo {pages} {868} (\bibinfo {year} {2022})}\BibitemShut {NoStop}%
\bibitem [{\citenamefont {Figueroa-Romero}\ \emph {et~al.}(2023)\citenamefont {Figueroa-Romero}, \citenamefont {Papi{\v{c}}}, \citenamefont {Auer}, \citenamefont {Hsieh}, \citenamefont {Modi},\ and\ \citenamefont {de~Vega}}]{figueroa2023operational}%
  \BibitemOpen
  \bibfield  {author} {\bibinfo {author} {\bibfnamefont {P.}~\bibnamefont {Figueroa-Romero}}, \bibinfo {author} {\bibfnamefont {M.}~\bibnamefont {Papi{\v{c}}}}, \bibinfo {author} {\bibfnamefont {A.}~\bibnamefont {Auer}}, \bibinfo {author} {\bibfnamefont {M.-H.}\ \bibnamefont {Hsieh}}, \bibinfo {author} {\bibfnamefont {K.}~\bibnamefont {Modi}},\ and\ \bibinfo {author} {\bibfnamefont {I.}~\bibnamefont {de~Vega}},\ }\bibfield  {journal} {\bibinfo  {journal} {arXiv preprint arXiv:2305.04704}\ }\href {https://doi.org/10.48550/arXiv.2305.04704} {10.48550/arXiv.2305.04704} (\bibinfo {year} {2023})\BibitemShut {NoStop}%
\bibitem [{\citenamefont {Milz}\ \emph {et~al.}(2018{\natexlab{a}})\citenamefont {Milz}, \citenamefont {Pollock},\ and\ \citenamefont {Modi}}]{milz2018reconstructing}%
  \BibitemOpen
  \bibfield  {author} {\bibinfo {author} {\bibfnamefont {S.}~\bibnamefont {Milz}}, \bibinfo {author} {\bibfnamefont {F.~A.}\ \bibnamefont {Pollock}},\ and\ \bibinfo {author} {\bibfnamefont {K.}~\bibnamefont {Modi}},\ }\href {https://doi.org/10.1103/PhysRevA.98.012108} {\bibfield  {journal} {\bibinfo  {journal} {Physical Review A}\ }\textbf {\bibinfo {volume} {98}},\ \bibinfo {pages} {012108} (\bibinfo {year} {2018}{\natexlab{a}})}\BibitemShut {NoStop}%
\bibitem [{\citenamefont {White}\ \emph {et~al.}(2020)\citenamefont {White}, \citenamefont {Hill}, \citenamefont {Pollock}, \citenamefont {Hollenberg},\ and\ \citenamefont {Modi}}]{white2020demonstration}%
  \BibitemOpen
  \bibfield  {author} {\bibinfo {author} {\bibfnamefont {G.~A.}\ \bibnamefont {White}}, \bibinfo {author} {\bibfnamefont {C.~D.}\ \bibnamefont {Hill}}, \bibinfo {author} {\bibfnamefont {F.~A.}\ \bibnamefont {Pollock}}, \bibinfo {author} {\bibfnamefont {L.~C.}\ \bibnamefont {Hollenberg}},\ and\ \bibinfo {author} {\bibfnamefont {K.}~\bibnamefont {Modi}},\ }\href {https://doi.org/10.1038/s41467-020-20113-3} {\bibfield  {journal} {\bibinfo  {journal} {Nature Communications}\ }\textbf {\bibinfo {volume} {11}},\ \bibinfo {pages} {6301} (\bibinfo {year} {2020})}\BibitemShut {NoStop}%
\bibitem [{\citenamefont {White}\ \emph {et~al.}(2021)\citenamefont {White}, \citenamefont {Pollock}, \citenamefont {Hollenberg}, \citenamefont {Hill},\ and\ \citenamefont {Modi}}]{white2021many}%
  \BibitemOpen
  \bibfield  {author} {\bibinfo {author} {\bibfnamefont {G.~A.}\ \bibnamefont {White}}, \bibinfo {author} {\bibfnamefont {F.~A.}\ \bibnamefont {Pollock}}, \bibinfo {author} {\bibfnamefont {L.~C.}\ \bibnamefont {Hollenberg}}, \bibinfo {author} {\bibfnamefont {C.~D.}\ \bibnamefont {Hill}},\ and\ \bibinfo {author} {\bibfnamefont {K.}~\bibnamefont {Modi}},\ }\bibfield  {journal} {\bibinfo  {journal} {arXiv preprint arXiv:2107.13934}\ }\href {https://doi.org/10.48550/arXiv.2107.13934} {10.48550/arXiv.2107.13934} (\bibinfo {year} {2021})\BibitemShut {NoStop}%
\bibitem [{\citenamefont {White}\ \emph {et~al.}(2022)\citenamefont {White}, \citenamefont {Pollock}, \citenamefont {Hollenberg}, \citenamefont {Modi},\ and\ \citenamefont {Hill}}]{white2022non}%
  \BibitemOpen
  \bibfield  {author} {\bibinfo {author} {\bibfnamefont {G.~A.}\ \bibnamefont {White}}, \bibinfo {author} {\bibfnamefont {F.~A.}\ \bibnamefont {Pollock}}, \bibinfo {author} {\bibfnamefont {L.~C.}\ \bibnamefont {Hollenberg}}, \bibinfo {author} {\bibfnamefont {K.}~\bibnamefont {Modi}},\ and\ \bibinfo {author} {\bibfnamefont {C.~D.}\ \bibnamefont {Hill}},\ }\href {https://doi.org/10.1103/PRXQuantum.3.020344} {\bibfield  {journal} {\bibinfo  {journal} {PRX Quantum}\ }\textbf {\bibinfo {volume} {3}},\ \bibinfo {pages} {020344} (\bibinfo {year} {2022})}\BibitemShut {NoStop}%
\bibitem [{\citenamefont {White}(2022)}]{white2022characterization}%
  \BibitemOpen
  \bibfield  {author} {\bibinfo {author} {\bibfnamefont {G.}~\bibnamefont {White}},\ }\href {https://doi.org/10.1038/s42254-022-00446-2} {\bibfield  {journal} {\bibinfo  {journal} {Nature Reviews Physics}\ }\textbf {\bibinfo {volume} {4}},\ \bibinfo {pages} {287} (\bibinfo {year} {2022})}\BibitemShut {NoStop}%
\bibitem [{\citenamefont {White}\ \emph {et~al.}(2023)\citenamefont {White}, \citenamefont {Modi},\ and\ \citenamefont {Hill}}]{white2023filtering}%
  \BibitemOpen
  \bibfield  {author} {\bibinfo {author} {\bibfnamefont {G.~A.}\ \bibnamefont {White}}, \bibinfo {author} {\bibfnamefont {K.}~\bibnamefont {Modi}},\ and\ \bibinfo {author} {\bibfnamefont {C.~D.}\ \bibnamefont {Hill}},\ }\href {https://doi.org/10.1103/PhysRevLett.130.160401} {\bibfield  {journal} {\bibinfo  {journal} {Physical Review Letters}\ }\textbf {\bibinfo {volume} {130}},\ \bibinfo {pages} {160401} (\bibinfo {year} {2023})}\BibitemShut {NoStop}%
\bibitem [{\citenamefont {Aloisio}\ \emph {et~al.}(2023)\citenamefont {Aloisio}, \citenamefont {White}, \citenamefont {Hill},\ and\ \citenamefont {Modi}}]{aloisio2023sampling}%
  \BibitemOpen
  \bibfield  {author} {\bibinfo {author} {\bibfnamefont {I.~A.}\ \bibnamefont {Aloisio}}, \bibinfo {author} {\bibfnamefont {G.~A.}\ \bibnamefont {White}}, \bibinfo {author} {\bibfnamefont {C.~D.}\ \bibnamefont {Hill}},\ and\ \bibinfo {author} {\bibfnamefont {K.}~\bibnamefont {Modi}},\ }\href {https://doi.org/10.1103/PRXQuantum.4.020310} {\bibfield  {journal} {\bibinfo  {journal} {PRX Quantum}\ }\textbf {\bibinfo {volume} {4}},\ \bibinfo {pages} {020310} (\bibinfo {year} {2023})}\BibitemShut {NoStop}%
\bibitem [{\citenamefont {Strasberg}(2019)}]{strasberg2019repeated}%
  \BibitemOpen
  \bibfield  {author} {\bibinfo {author} {\bibfnamefont {P.}~\bibnamefont {Strasberg}},\ }\href {https://doi.org/10.1103/PhysRevLett.123.180604} {\bibfield  {journal} {\bibinfo  {journal} {Physical review letters}\ }\textbf {\bibinfo {volume} {123}},\ \bibinfo {pages} {180604} (\bibinfo {year} {2019})}\BibitemShut {NoStop}%
\bibitem [{\citenamefont {Figueroa-Romero}\ \emph {et~al.}(2020)\citenamefont {Figueroa-Romero}, \citenamefont {Modi},\ and\ \citenamefont {Pollock}}]{figueroa2020equilibration}%
  \BibitemOpen
  \bibfield  {author} {\bibinfo {author} {\bibfnamefont {P.}~\bibnamefont {Figueroa-Romero}}, \bibinfo {author} {\bibfnamefont {K.}~\bibnamefont {Modi}},\ and\ \bibinfo {author} {\bibfnamefont {F.~A.}\ \bibnamefont {Pollock}},\ }\href {https://doi.org/10.1103/PhysRevE.102.032144} {\bibfield  {journal} {\bibinfo  {journal} {Physical Review E}\ }\textbf {\bibinfo {volume} {102}},\ \bibinfo {pages} {032144} (\bibinfo {year} {2020})}\BibitemShut {NoStop}%
\bibitem [{\citenamefont {Huang}(2022)}]{huang2022fluctuation}%
  \BibitemOpen
  \bibfield  {author} {\bibinfo {author} {\bibfnamefont {Z.}~\bibnamefont {Huang}},\ }\href {https://doi.org/10.1103/PhysRevA.105.062217} {\bibfield  {journal} {\bibinfo  {journal} {Phys. Rev. A}\ }\textbf {\bibinfo {volume} {105}},\ \bibinfo {pages} {062217} (\bibinfo {year} {2022})}\BibitemShut {NoStop}%
\bibitem [{\citenamefont {Huang}(2023)}]{huang2023multiple}%
  \BibitemOpen
  \bibfield  {author} {\bibinfo {author} {\bibfnamefont {Z.}~\bibnamefont {Huang}},\ }\href {https://doi.org/10.1103/PhysRevA.108.032217} {\bibfield  {journal} {\bibinfo  {journal} {Phys. Rev. A}\ }\textbf {\bibinfo {volume} {108}},\ \bibinfo {pages} {032217} (\bibinfo {year} {2023})}\BibitemShut {NoStop}%
\bibitem [{\citenamefont {Dowling}\ \emph {et~al.}(2023{\natexlab{a}})\citenamefont {Dowling}, \citenamefont {Figueroa-Romero}, \citenamefont {Pollock}, \citenamefont {Strasberg},\ and\ \citenamefont {Modi}}]{dowling2023relaxation}%
  \BibitemOpen
  \bibfield  {author} {\bibinfo {author} {\bibfnamefont {N.}~\bibnamefont {Dowling}}, \bibinfo {author} {\bibfnamefont {P.}~\bibnamefont {Figueroa-Romero}}, \bibinfo {author} {\bibfnamefont {F.~A.}\ \bibnamefont {Pollock}}, \bibinfo {author} {\bibfnamefont {P.}~\bibnamefont {Strasberg}},\ and\ \bibinfo {author} {\bibfnamefont {K.}~\bibnamefont {Modi}},\ }\href {https://doi.org/10.22331/q-2023-06-01-1027} {\bibfield  {journal} {\bibinfo  {journal} {Quantum}\ }\textbf {\bibinfo {volume} {7}},\ \bibinfo {pages} {1027} (\bibinfo {year} {2023}{\natexlab{a}})}\BibitemShut {NoStop}%
\bibitem [{\citenamefont {Dowling}\ \emph {et~al.}(2023{\natexlab{b}})\citenamefont {Dowling}, \citenamefont {Figueroa-Romero}, \citenamefont {Pollock}, \citenamefont {Strasberg},\ and\ \citenamefont {Modi}}]{dowling2023equilibration}%
  \BibitemOpen
  \bibfield  {author} {\bibinfo {author} {\bibfnamefont {N.}~\bibnamefont {Dowling}}, \bibinfo {author} {\bibfnamefont {P.}~\bibnamefont {Figueroa-Romero}}, \bibinfo {author} {\bibfnamefont {F.~A.}\ \bibnamefont {Pollock}}, \bibinfo {author} {\bibfnamefont {P.}~\bibnamefont {Strasberg}},\ and\ \bibinfo {author} {\bibfnamefont {K.}~\bibnamefont {Modi}},\ }\href {https://doi.org/10.21468/SciPostPhysCore.6.2.043} {\bibfield  {journal} {\bibinfo  {journal} {SciPost Physics Core}\ }\textbf {\bibinfo {volume} {6}},\ \bibinfo {pages} {043} (\bibinfo {year} {2023}{\natexlab{b}})}\BibitemShut {NoStop}%
\bibitem [{\citenamefont {Guo}\ \emph {et~al.}(2020)\citenamefont {Guo}, \citenamefont {Modi},\ and\ \citenamefont {Poletti}}]{guo2020tensor}%
  \BibitemOpen
  \bibfield  {author} {\bibinfo {author} {\bibfnamefont {C.}~\bibnamefont {Guo}}, \bibinfo {author} {\bibfnamefont {K.}~\bibnamefont {Modi}},\ and\ \bibinfo {author} {\bibfnamefont {D.}~\bibnamefont {Poletti}},\ }\href {https://doi.org/10.1103/PhysRevA.102.062414} {\bibfield  {journal} {\bibinfo  {journal} {Phys. Rev. A}\ }\textbf {\bibinfo {volume} {102}},\ \bibinfo {pages} {062414} (\bibinfo {year} {2020})}\BibitemShut {NoStop}%
\bibitem [{\citenamefont {Huang}\ and\ \citenamefont {Guo}(2023)}]{huang2023leggettgarg}%
  \BibitemOpen
  \bibfield  {author} {\bibinfo {author} {\bibfnamefont {Z.}~\bibnamefont {Huang}}\ and\ \bibinfo {author} {\bibfnamefont {X.-K.}\ \bibnamefont {Guo}},\ }\bibfield  {journal} {\bibinfo  {journal} {arXiv preprint arXiv:2211.13396}\ }\href {https://doi.org/10.48550/arXiv.2211.13396} {10.48550/arXiv.2211.13396} (\bibinfo {year} {2023})\BibitemShut {NoStop}%
\bibitem [{\citenamefont {Butler}\ \emph {et~al.}(2023)\citenamefont {Butler}, \citenamefont {Fux}, \citenamefont {Lovett}, \citenamefont {Keeling},\ and\ \citenamefont {Eastham}}]{butler2023optimizing}%
  \BibitemOpen
  \bibfield  {author} {\bibinfo {author} {\bibfnamefont {E.~P.}\ \bibnamefont {Butler}}, \bibinfo {author} {\bibfnamefont {G.}~\bibnamefont {Fux}}, \bibinfo {author} {\bibfnamefont {B.~W.}\ \bibnamefont {Lovett}}, \bibinfo {author} {\bibfnamefont {J.}~\bibnamefont {Keeling}},\ and\ \bibinfo {author} {\bibfnamefont {P.~R.}\ \bibnamefont {Eastham}},\ }\bibfield  {journal} {\bibinfo  {journal} {arXiv preprint arXiv:2303.16002}\ }\href {https://doi.org/10.48550/arXiv.2303.16002} {10.48550/arXiv.2303.16002} (\bibinfo {year} {2023})\BibitemShut {NoStop}%
\bibitem [{\citenamefont {Chiribella}\ \emph {et~al.}(2008)\citenamefont {Chiribella}, \citenamefont {D'Ariano},\ and\ \citenamefont {Perinotti}}]{chribella2008quantum}%
  \BibitemOpen
  \bibfield  {author} {\bibinfo {author} {\bibfnamefont {G.}~\bibnamefont {Chiribella}}, \bibinfo {author} {\bibfnamefont {G.~M.}\ \bibnamefont {D'Ariano}},\ and\ \bibinfo {author} {\bibfnamefont {P.}~\bibnamefont {Perinotti}},\ }\href {https://doi.org/10.1103/PhysRevLett.101.060401} {\bibfield  {journal} {\bibinfo  {journal} {Phys. Rev. Lett.}\ }\textbf {\bibinfo {volume} {101}},\ \bibinfo {pages} {060401} (\bibinfo {year} {2008})}\BibitemShut {NoStop}%
\bibitem [{\citenamefont {Chiribella}\ \emph {et~al.}(2009)\citenamefont {Chiribella}, \citenamefont {D’Ariano},\ and\ \citenamefont {Perinotti}}]{chiribella2009theoretical}%
  \BibitemOpen
  \bibfield  {author} {\bibinfo {author} {\bibfnamefont {G.}~\bibnamefont {Chiribella}}, \bibinfo {author} {\bibfnamefont {G.~M.}\ \bibnamefont {D’Ariano}},\ and\ \bibinfo {author} {\bibfnamefont {P.}~\bibnamefont {Perinotti}},\ }\href {https://doi.org/10.1103/PhysRevA.80.022339} {\bibfield  {journal} {\bibinfo  {journal} {Physical Review A}\ }\textbf {\bibinfo {volume} {80}},\ \bibinfo {pages} {022339} (\bibinfo {year} {2009})}\BibitemShut {NoStop}%
\bibitem [{\citenamefont {Costa}\ and\ \citenamefont {Shrapnel}(2016)}]{costa2016quantum}%
  \BibitemOpen
  \bibfield  {author} {\bibinfo {author} {\bibfnamefont {F.}~\bibnamefont {Costa}}\ and\ \bibinfo {author} {\bibfnamefont {S.}~\bibnamefont {Shrapnel}},\ }\href {https://doi.org/10.1088/1367-2630/18/6/063032} {\bibfield  {journal} {\bibinfo  {journal} {New Journal of Physics}\ }\textbf {\bibinfo {volume} {18}},\ \bibinfo {pages} {063032} (\bibinfo {year} {2016})}\BibitemShut {NoStop}%
\bibitem [{\citenamefont {Costa}\ \emph {et~al.}(2018)\citenamefont {Costa}, \citenamefont {Ringbauer}, \citenamefont {Goggin}, \citenamefont {White},\ and\ \citenamefont {Fedrizzi}}]{costa2018unifying}%
  \BibitemOpen
  \bibfield  {author} {\bibinfo {author} {\bibfnamefont {F.}~\bibnamefont {Costa}}, \bibinfo {author} {\bibfnamefont {M.}~\bibnamefont {Ringbauer}}, \bibinfo {author} {\bibfnamefont {M.~E.}\ \bibnamefont {Goggin}}, \bibinfo {author} {\bibfnamefont {A.~G.}\ \bibnamefont {White}},\ and\ \bibinfo {author} {\bibfnamefont {A.}~\bibnamefont {Fedrizzi}},\ }\href {https://doi.org/10.1103/PhysRevA.98.012328} {\bibfield  {journal} {\bibinfo  {journal} {Phys. Rev. A}\ }\textbf {\bibinfo {volume} {98}},\ \bibinfo {pages} {012328} (\bibinfo {year} {2018})}\BibitemShut {NoStop}%
\bibitem [{\citenamefont {Milz}\ \emph {et~al.}(2018{\natexlab{b}})\citenamefont {Milz}, \citenamefont {Pollock}, \citenamefont {Le}, \citenamefont {Chiribella},\ and\ \citenamefont {Modi}}]{milz2018entanglement}%
  \BibitemOpen
  \bibfield  {author} {\bibinfo {author} {\bibfnamefont {S.}~\bibnamefont {Milz}}, \bibinfo {author} {\bibfnamefont {F.~A.}\ \bibnamefont {Pollock}}, \bibinfo {author} {\bibfnamefont {T.~P.}\ \bibnamefont {Le}}, \bibinfo {author} {\bibfnamefont {G.}~\bibnamefont {Chiribella}},\ and\ \bibinfo {author} {\bibfnamefont {K.}~\bibnamefont {Modi}},\ }\href {https://doi.org/10.1088/1367-2630/aaafee} {\bibfield  {journal} {\bibinfo  {journal} {New Journal of Physics}\ }\textbf {\bibinfo {volume} {20}},\ \bibinfo {pages} {033033} (\bibinfo {year} {2018}{\natexlab{b}})}\BibitemShut {NoStop}%
\bibitem [{\citenamefont {Nery}\ \emph {et~al.}(2021)\citenamefont {Nery}, \citenamefont {Quintino}, \citenamefont {Gu{\'{e}}rin}, \citenamefont {Maciel},\ and\ \citenamefont {Vianna}}]{nery2021simple}%
  \BibitemOpen
  \bibfield  {author} {\bibinfo {author} {\bibfnamefont {M.}~\bibnamefont {Nery}}, \bibinfo {author} {\bibfnamefont {M.~T.}\ \bibnamefont {Quintino}}, \bibinfo {author} {\bibfnamefont {P.~A.}\ \bibnamefont {Gu{\'{e}}rin}}, \bibinfo {author} {\bibfnamefont {T.~O.}\ \bibnamefont {Maciel}},\ and\ \bibinfo {author} {\bibfnamefont {R.~O.}\ \bibnamefont {Vianna}},\ }\href {https://doi.org/10.22331/q-2021-09-09-538} {\bibfield  {journal} {\bibinfo  {journal} {{Quantum}}\ }\textbf {\bibinfo {volume} {5}},\ \bibinfo {pages} {538} (\bibinfo {year} {2021})}\BibitemShut {NoStop}%
\bibitem [{\citenamefont {Giarmatzi}\ and\ \citenamefont {Costa}(2021)}]{giarmatzi2021witnessing}%
  \BibitemOpen
  \bibfield  {author} {\bibinfo {author} {\bibfnamefont {C.}~\bibnamefont {Giarmatzi}}\ and\ \bibinfo {author} {\bibfnamefont {F.}~\bibnamefont {Costa}},\ }\href {https://doi.org/10.22331/q-2021-04-26-440} {\bibfield  {journal} {\bibinfo  {journal} {{Quantum}}\ }\textbf {\bibinfo {volume} {5}},\ \bibinfo {pages} {440} (\bibinfo {year} {2021})}\BibitemShut {NoStop}%
\bibitem [{\citenamefont {Milz}\ \emph {et~al.}(2022)\citenamefont {Milz}, \citenamefont {Bavaresco},\ and\ \citenamefont {Chiribella}}]{milz2022resource}%
  \BibitemOpen
  \bibfield  {author} {\bibinfo {author} {\bibfnamefont {S.}~\bibnamefont {Milz}}, \bibinfo {author} {\bibfnamefont {J.}~\bibnamefont {Bavaresco}},\ and\ \bibinfo {author} {\bibfnamefont {G.}~\bibnamefont {Chiribella}},\ }\href {https://doi.org/10.22331/q-2022-08-25-788} {\bibfield  {journal} {\bibinfo  {journal} {Quantum}\ }\textbf {\bibinfo {volume} {6}},\ \bibinfo {pages} {788} (\bibinfo {year} {2022})}\BibitemShut {NoStop}%
\bibitem [{\citenamefont {Milz}\ and\ \citenamefont {Quintino}(2023)}]{milz2023transformations}%
  \BibitemOpen
  \bibfield  {author} {\bibinfo {author} {\bibfnamefont {S.}~\bibnamefont {Milz}}\ and\ \bibinfo {author} {\bibfnamefont {M.~T.}\ \bibnamefont {Quintino}},\ }\bibfield  {journal} {\bibinfo  {journal} {arXiv preprint arXiv:2305.01247}\ }\href {https://doi.org/10.48550/arXiv.2305.01247} {10.48550/arXiv.2305.01247} (\bibinfo {year} {2023})\BibitemShut {NoStop}%
\bibitem [{\citenamefont {Berk}\ \emph {et~al.}(2021)\citenamefont {Berk}, \citenamefont {Garner}, \citenamefont {Yadin}, \citenamefont {Modi},\ and\ \citenamefont {Pollock}}]{berk2021resource}%
  \BibitemOpen
  \bibfield  {author} {\bibinfo {author} {\bibfnamefont {G.~D.}\ \bibnamefont {Berk}}, \bibinfo {author} {\bibfnamefont {A.~J.}\ \bibnamefont {Garner}}, \bibinfo {author} {\bibfnamefont {B.}~\bibnamefont {Yadin}}, \bibinfo {author} {\bibfnamefont {K.}~\bibnamefont {Modi}},\ and\ \bibinfo {author} {\bibfnamefont {F.~A.}\ \bibnamefont {Pollock}},\ }\href {https://doi.org/10.22331/q-2021-04-20-435} {\bibfield  {journal} {\bibinfo  {journal} {Quantum}\ }\textbf {\bibinfo {volume} {5}},\ \bibinfo {pages} {435} (\bibinfo {year} {2021})}\BibitemShut {NoStop}%
\bibitem [{\citenamefont {Berk}\ \emph {et~al.}(2023)\citenamefont {Berk}, \citenamefont {Milz}, \citenamefont {Pollock},\ and\ \citenamefont {Modi}}]{berk2023extracting}%
  \BibitemOpen
  \bibfield  {author} {\bibinfo {author} {\bibfnamefont {G.~D.}\ \bibnamefont {Berk}}, \bibinfo {author} {\bibfnamefont {S.}~\bibnamefont {Milz}}, \bibinfo {author} {\bibfnamefont {F.~A.}\ \bibnamefont {Pollock}},\ and\ \bibinfo {author} {\bibfnamefont {K.}~\bibnamefont {Modi}},\ }\href {https://doi.org/10.1038/s41534-023-00774-w} {\bibfield  {journal} {\bibinfo  {journal} {npj Quantum Information}\ }\textbf {\bibinfo {volume} {9}},\ \bibinfo {pages} {104} (\bibinfo {year} {2023})}\BibitemShut {NoStop}%
\bibitem [{\citenamefont {Chitambar}\ and\ \citenamefont {Gour}(2019)}]{chitambar2019quantum}%
  \BibitemOpen
  \bibfield  {author} {\bibinfo {author} {\bibfnamefont {E.}~\bibnamefont {Chitambar}}\ and\ \bibinfo {author} {\bibfnamefont {G.}~\bibnamefont {Gour}},\ }\href {https://doi.org/10.1103/RevModPhys.91.025001} {\bibfield  {journal} {\bibinfo  {journal} {Rev. Mod. Phys.}\ }\textbf {\bibinfo {volume} {91}},\ \bibinfo {pages} {025001} (\bibinfo {year} {2019})}\BibitemShut {NoStop}%
\bibitem [{\citenamefont {Luo}\ \emph {et~al.}(2012)\citenamefont {Luo}, \citenamefont {Fu},\ and\ \citenamefont {Song}}]{luo2012quantifying}%
  \BibitemOpen
  \bibfield  {author} {\bibinfo {author} {\bibfnamefont {S.}~\bibnamefont {Luo}}, \bibinfo {author} {\bibfnamefont {S.}~\bibnamefont {Fu}},\ and\ \bibinfo {author} {\bibfnamefont {H.}~\bibnamefont {Song}},\ }\href {https://doi.org/10.1103/PhysRevA.86.044101} {\bibfield  {journal} {\bibinfo  {journal} {Phys. Rev. A}\ }\textbf {\bibinfo {volume} {86}},\ \bibinfo {pages} {044101} (\bibinfo {year} {2012})}\BibitemShut {NoStop}%
\bibitem [{\citenamefont {Bylicka}\ \emph {et~al.}(2014)\citenamefont {Bylicka}, \citenamefont {Chru{\'s}ci{\'n}ski},\ and\ \citenamefont {Maniscalco}}]{bylicka2014non}%
  \BibitemOpen
  \bibfield  {author} {\bibinfo {author} {\bibfnamefont {B.}~\bibnamefont {Bylicka}}, \bibinfo {author} {\bibfnamefont {D.}~\bibnamefont {Chru{\'s}ci{\'n}ski}},\ and\ \bibinfo {author} {\bibfnamefont {S.}~\bibnamefont {Maniscalco}},\ }\href {https://doi.org/10.1038/srep05720} {\bibfield  {journal} {\bibinfo  {journal} {Scientific reports}\ }\textbf {\bibinfo {volume} {4}},\ \bibinfo {pages} {5720} (\bibinfo {year} {2014})}\BibitemShut {NoStop}%
\bibitem [{\citenamefont {Bylicka}\ \emph {et~al.}(2016)\citenamefont {Bylicka}, \citenamefont {Tukiainen}, \citenamefont {Chru{\'s}ci{\'n}ski}, \citenamefont {Piilo},\ and\ \citenamefont {Maniscalco}}]{bylicka2016thermodynamic}%
  \BibitemOpen
  \bibfield  {author} {\bibinfo {author} {\bibfnamefont {B.}~\bibnamefont {Bylicka}}, \bibinfo {author} {\bibfnamefont {M.}~\bibnamefont {Tukiainen}}, \bibinfo {author} {\bibfnamefont {D.}~\bibnamefont {Chru{\'s}ci{\'n}ski}}, \bibinfo {author} {\bibfnamefont {J.}~\bibnamefont {Piilo}},\ and\ \bibinfo {author} {\bibfnamefont {S.}~\bibnamefont {Maniscalco}},\ }\href {https://doi.org/10.1038/srep27989} {\bibfield  {journal} {\bibinfo  {journal} {Scientific reports}\ }\textbf {\bibinfo {volume} {6}},\ \bibinfo {pages} {27989} (\bibinfo {year} {2016})}\BibitemShut {NoStop}%
\bibitem [{\citenamefont {Verstraete}\ and\ \citenamefont {Verschelde}(2002)}]{verstraete2002quantum}%
  \BibitemOpen
  \bibfield  {author} {\bibinfo {author} {\bibfnamefont {F.}~\bibnamefont {Verstraete}}\ and\ \bibinfo {author} {\bibfnamefont {H.}~\bibnamefont {Verschelde}},\ }\bibfield  {journal} {\bibinfo  {journal} {arXiv preprint quant-ph/0202124}\ }\href {https://doi.org/10.48550/arXiv.quant-ph/0202124} {10.48550/arXiv.quant-ph/0202124} (\bibinfo {year} {2002})\BibitemShut {NoStop}%
\bibitem [{\citenamefont {Stinespring}(1955)}]{stinespring1955positive}%
  \BibitemOpen
  \bibfield  {author} {\bibinfo {author} {\bibfnamefont {W.~F.}\ \bibnamefont {Stinespring}},\ }\href {https://doi.org/10.2307/2032342} {\bibfield  {journal} {\bibinfo  {journal} {Proceedings of the American Mathematical Society}\ }\textbf {\bibinfo {volume} {6}},\ \bibinfo {pages} {211} (\bibinfo {year} {1955})}\BibitemShut {NoStop}%
\bibitem [{\citenamefont {Araki}\ and\ \citenamefont {Lieb}(1970)}]{araki1970entropy}%
  \BibitemOpen
  \bibfield  {author} {\bibinfo {author} {\bibfnamefont {H.}~\bibnamefont {Araki}}\ and\ \bibinfo {author} {\bibfnamefont {E.~H.}\ \bibnamefont {Lieb}},\ }\href {https://doi.org/10.1007/BF01646092} {\bibfield  {journal} {\bibinfo  {journal} {Communications in Mathematical Physics}\ }\textbf {\bibinfo {volume} {18}},\ \bibinfo {pages} {160} (\bibinfo {year} {1970})}\BibitemShut {NoStop}%
\bibitem [{\citenamefont {Faist}\ \emph {et~al.}(2019)\citenamefont {Faist}, \citenamefont {Berta},\ and\ \citenamefont {Brand{\~a}o}}]{faist2019thermodynamic}%
  \BibitemOpen
  \bibfield  {author} {\bibinfo {author} {\bibfnamefont {P.}~\bibnamefont {Faist}}, \bibinfo {author} {\bibfnamefont {M.}~\bibnamefont {Berta}},\ and\ \bibinfo {author} {\bibfnamefont {F.}~\bibnamefont {Brand{\~a}o}},\ }\href@noop {} {\bibfield  {journal} {\bibinfo  {journal} {Physical review letters}\ }\textbf {\bibinfo {volume} {122}},\ \bibinfo {pages} {200601} (\bibinfo {year} {2019})}\BibitemShut {NoStop}%
\end{thebibliography}

%

\end{document}